\documentclass[letterpaper,twocolumn,10pt]{article}
\pdfoutput=1
\usepackage{usenix-2020-09}
\usepackage{tikz}
\usepackage{amsmath}

\usepackage{filecontents}
\usepackage{setspace}
\usepackage{setspace}
\usepackage{booktabs} % For formal tables
\usepackage{soul,xcolor}
\usepackage{listings}
\usepackage{amsmath}
\usepackage{color}
\usepackage{xcolor}
\usepackage{lipsum}
\usepackage{comment}
\usepackage[title]{appendix}
\usepackage{multirow}
\usepackage{booktabs}
\usepackage{xspace}
\usepackage{graphicx}
\usepackage{caption}
\usepackage{pdfpages}
\usepackage{setspace}
\usepackage{subfiles}
\usepackage{booktabs}
\usepackage{array}
\usepackage{enumitem}
\usepackage{subcaption}
\usepackage{pifont}
\usepackage[font=small,labelfont=bf]{caption}
\usepackage{url}
\newcommand{\squishend}{
    \end{list}  }

\definecolor{dong}{RGB}{0,0,255} %{0,0,0}%{255,0,0} 
\definecolor{dong2}{RGB}{0,0,0}%{0,0,0}%{0,0,255}
\definecolor{jie}{rgb}{0.0, 0.5, 0.0}
\definecolor{check}{RGB}{0,0,0}

\renewcommand{\baselinestretch}{1.0} 
\newcommand{\name}{ZeRO-Offload\xspace}

\setlist{nolistsep}

\begin{document}
%-------------------------------------------------------------------------------
%don't want date printed
%\date{}
% \title{\Large \bf \name: Democratizing Large-Scale Model Training}

\title{\Large \bf \name: Democratizing Billion-Scale Model Training}

%for single author (just remove % characters)
\begin{comment}
\author{
{%\rm Your N.\ Here
}\\
%Your Institution
\and
{%\rm Second Name
}\\
%Second Institution
% copy the following lines to add more authors
% \and
% {\rm Name}\\
%Name Institution
} % end author
\end{comment}

\author{Jie Ren$^\ast$
, Samyam Rajbhandari$^\dagger$
, Reza Yazdani Aminabadi$^\dagger$
, Olatunji Ruwase$^\dagger$
\\ Shuangyan Yang$^\ast$
, Minjia Zhang$^\dagger$
, Dong Li$^\ast$
, Yuxiong He$^\dagger$
\\
%Your Institution
$^\dagger$Microsoft ,   
$^\ast$University of California, Merced
\\
\{jren6, syang127, dli35\}@ucmerced.edu, \{samyamr, yazdani.reza, olruwase, minjiaz, yuxhe\}@microsoft.com
\and
{%\rm Second Name
}\\
%Second Institution
% copy the following lines to add more authors
% \and
% {\rm Name}\\
%Name Institution
} % end author
% \end{comment}
\maketitle

%-------------------------------------------------------------------------------
\begin{abstract}
%-------------------------------------------------------------------------------
Large-scale model training has been a playing ground for a limited few requiring complex model refactoring and access to prohibitively expensive GPU clusters. \name changes the large model training landscape by making large model training accessible to nearly everyone. It can train models with over 13 billion parameters on a single GPU, a 10x increase in size compared to popular framework such as PyTorch, and it does so without requiring any model change from the data scientists or sacrificing computational efficiency.

\name enables large model training by offloading data and compute to CPU. To preserve compute efficiency, it is designed to minimize the data movement to/from GPU, and reduce CPU compute time while maximizing memory savings on GPU. As a result, \name can achieve 40 TFlops/GPU on a single NVIDIA V100 GPU for 10B parameter model compared to 30TF using PyTorch alone for a 1.4B parameter model, the largest that can be trained without running out of memory. \name is also designed to scale on multiple-GPUs when available, offering near linear speedup on up to 128 GPUs. Additionally, it can work together with model parallelism to train models with over 70 billion parameters on a single DGX-2 box, a 4.5x increase in model size compared to using model parallelism alone.

By combining compute and memory efficiency with ease-of-use, \name democratizes large-scale model training making it accessible to even data scientists with access to just a single GPU.
\end{abstract}

%-------------------------------------------------------------------------------

% \input text/intro 
% \input text/background    
% \input text/design
% \input text/scheduling-and-mapping
% \input text/optimization
% \input text/evaluation
% %\input text/related_work  
% \input text/conclusions 
\section{Introduction}
\label{s:intro}
Since the advent of the attention-based deep learning (DL) models in 2017, we have seen an exponential growth in DL model size, fueled by substantial quality gains that these attention based models can offer with the increase in the number of parameters. For example, the largest language model in literature had less than 100M parameters in 2017, it grew to over 300M with BERT \cite{devlin2018bert} in 2018, increasing to tens of billions in 2019 with models such as GPT-2 \cite{cao2020pretrained}, T5 \cite{raffel2020exploring}, Megatron-LM \cite{Megatron-LM} and Turing-NLG \cite{turing-nlg}. Today, the largest language model GPT-3 \cite {brown2020language} has a staggering number of 175B parameters. With the three orders of magnitude growth in model size since 2017, the model accuracy continues to improve with the model size \cite{open-aiscaling}. Recent studies in fact show that larger models are more resource-efficient to train than smaller ones \cite{open-aiscaling} for a given accuracy target. As a result, we expect the model size to continue growing in the future.

%Pre-training large language models such as the ones described above requires non-trivial amount of computational resources, which is beyond the reach of many. However, once these models are pre-trained, fine-tuning these models to specific tasks requires significantly less compute resource that is available to many. For example, pre-training the Bert-Large model takes several hours on a 64-GPU NVIDIA V100 cluster \cite{fastestBERT}, while fine-tuning it for the Squad \cite{bert_github} takes less than an hour on a single V100 GPU. Furthermore, recent work \cite{GShard} has shown that it is possible to leverage the power of large models with sub-linear compute cost that would make large model training possible with limited computational resources, making it accessible to many without access to massive supercomputers. 

%Despite this possibility, 
However, accessibility to large model training is severely limited by the nature of state-of-art system technologies. Those technologies make entry into the large model training space prohibitively expensive. To be more specific, distributed parallel DL training technologies such as pipeline parallelism \cite{huang2018gpipe}, model parallelism \cite{Megatron-LM}, and ZeRO \cite{zero} (Zero Redundancy Optimizer) allow transcending the memory boundaries of single GPU/accelerator device by splitting the model states (parameters, gradients and optimizer states) across multiple GPU devices, enabling massive models that would otherwise simply not fit in a single GPU memory. All record-breaking large models such as GPT-2, Megatron-LM, Turing-NLG, GPT-3, were trained using a combination of the aforementioned technologies. However, all of these DL parallel technologies require having enough GPU devices such that the aggregated GPU memory can hold the model states required for the training. For example, training a 10B parameter model efficiently requires a DGX-2 equivalent node with 16 NVIDIA V100 cards, which cost over $100K$, beyond the reach of many data scientists, and even many academic and industrial institutions.

\textbf {Heterogeneous DL training}  is a promising approach to reduce GPU memory requirement by exploiting CPU memory. Many efforts have been made in this direction \cite{vDNN_micro16,SuperNeurons_ppopp18,Layrub,SwapAdvisor_asplos20,AutoSawp,AutoTM_asplos20,Capuchin_asplos20,hpca21:ren,isca20:buddy_compression, hpca21:ren}. %However, nearly all of them focus on relatively small models (less than 500M), where the primary memory bottleneck is the activation memory; The model states (parameters, gradients, and optimizer states) only require a very small portion of the total memory \cite{zero, Capuchin_asplos20}.  On the contrary, the primary memory bottleneck for large model training is the memory states as opposed to the activations (see Sec.~\ref{s:model_states}), and there is an absence in literature studying these workloads for heterogeneous DL training.
Nearly all of them target CNN based models, where activation memory is the memory bottleneck, and model size is fairly small (less than 500M). However, the primary memory bottleneck for recent attention based large model training are the model states, instead of activation memory. There is an absence in literature studying these workloads for heterogeneous DL training. Additionally, existing efforts on heterogeneous training are further limited in two major ways: i) nearly all of them exploit CPU memory, but not CPU compute, which we show can be used to significantly reduce the CPU-GPU communication overhead, and ii) they are mostly designed for and evaluated on single GPU ~\cite{Layrub, SwapAdvisor_asplos20, hpca21:ren, AutoSawp}, without a clear path to scaling efficiently on multiple GPUs that is crucial for large model training.

Addressing the aforementioned limitation, we attempt to democratize large model training by developing \name, a novel heterogeneous DL training technology designed specifically for large model training. \name exploits both CPU memory and compute for offloading, while offering a clear path towards efficiently scaling on multiple GPUs by working with ZeRO-powered data parallelism \cite{zero}. Additionally, our first principle analysis shows that \name provides an optimal and the only optimal solution in maximizing memory saving while minimizing communication overhead and CPU compute overhead for large model training. 

\name is designed around three main pillars: i) Efficiency, ii) Scalabilty, and iii) Usability.

\textbf{Efficiency}: The offload strategy is designed with the goal of achieving comparable compute efficiency to the state-of-art non-offload strategies but for significantly larger models. To achieve this goal, we rely on first principle analysis to identify a \emph{unique optimal} computation and data partitioning strategy between CPU and GPU devices that is optimal in three key aspects: i) it requires orders-of-magnitude fewer computation on CPU compared to GPU, preventing the CPU compute from becoming a performance bottleneck, ii) it minimizes the communication volume between CPU and GPU preventing communication from being a bottleneck, and iii) it provably maximizes  memory savings on GPU while achieving minimum communication volume. 

Our analysis shows that to be optimal in the aforementioned regards, we must offload the gradients, optimizer states and  optimizer computation to CPU, while keeping the parameters and  forward and backward computation on GPU. This strategy enables a 10x increase in model size, with minimum communication and limited CPU computation, which allows us to train 13B parameters on a single NVIDIA V100 GPU at 40 
TFLOPS, compared to 30 TFLOPS on the same GPU with 1.2B parameters, the largest model that can be trained without any CPU offloading. 

Offloading optimizer computation requires CPU to perform $O(M)$ computation compared to $O(MB)$ on GPU where $M$ and $B$ are the model size and batch sizes. In most cases, the batch size is large, and CPU computation is not a bottleneck, but for small batch sizes, the CPU compute can be a bottleneck. We address this using two optimizations: i) an efficient \emph{CPU optimizer} that is up to 6x faster than the-state-of-art , and ii)One-step \emph{delayed parameter update} that allows overlapping the CPU optimizer step with GPU compute, while preserving accuracy. Together, they preserve efficiency for \name even with small batch sizes.

\textbf{Scalability}:  Good scalability is crucial to take advantage of multiple GPUs that may be available to some data scientists. In the DL community, data parallelism is generally used as the de facto standard to scale DL training to multiple GPUs \cite{NIPS2012_4687, parallel_sgd, jmlr20_data_parallelism}. However, it is not designed to work with heterogeneous training and presents scalability challenges because of the replication of data and computation in data parallel training. Data parallel training replicates all the model states such as optimizer states, parameters, and gradients, and it also replicates the optimizer computation on each GPU. Therefore, offloading model states or optimizer computation to CPU in combination with data parallelism will result in significant replication of communication and CPU compute: increase the CPU memory requirement proportionally to the data parallelism degree while limiting throughput scalability due to the increased communication.

\begin{comment}
in literature that can offload model states \cite{vDNN_micro16,,Layrub,SwapAdvisor_asplos20,AutoSawp,AutoTM_asplos20,hpca21:ren,isca20:buddy_compression}.

For example, training a model with 16-way data parallelism on a single DGX-2 box combined with \name would cause each data parallel process to store a copy of all the optimizer states on the shared CPU memory, while also performing optimizer computation on the shared CPU resources, resulting in 16x redundancy in CPU memory storage, CPU-GPU communication and CPU computation. 
  %and in fact, we are unaware of any CPU offload strategy that offers solutions on how to combine CPU offload with multi-GPU training effectively.
\end{comment}
To address these challenges, \name combines our offload strategy with ZeRO \cite{zero} powered data parallelism instead of traditional data parallelism. The symbiosis allows \name to maintain a single copy of the optimizer states on the CPU memory regardless of the data parallel degree. Furthermore, it keeps the aggregate communication volume between GPU and CPU, as well as the aggregate CPU computation a constant regardless of data parallelism, allowing  \name to effectively utilize the linear increase in CPU compute with the increase in the data parallelism degree. As a result, \name achieves excellent scalability on up to 128 GPUs.

In addition to working with ZeRO powered data parallelism, \name can be combined with model parallelism \cite{mesh-tensorflow, Megatron-LM} to achieve higher memory savings, when multiple GPUs are available.  

\begin{figure}[!tbp]
  \centering
   \includegraphics[ width=\columnwidth, height=1.5in]{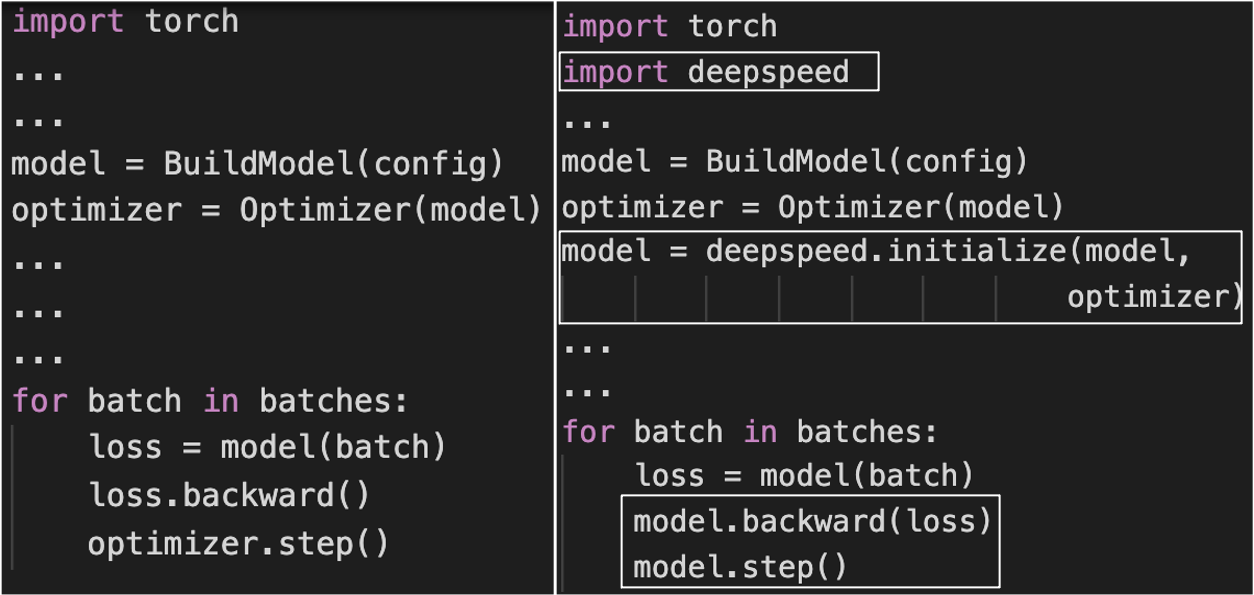}
   \vspace{-20pt}
  \caption{\name can be enabled with a few lines of change. The code on left shows a standard training pipeline, while the right shows the same pipeline with \name enabled.}
  \vspace{-10pt} 
\label{fig:ease-of-use}
  \vspace{-5pt} 
\end{figure}

\textbf{Usability}:  \name is available as part of an OpenSource PyTorch library, DeepSpeed (\url{www.deepspeed.ai}). Unlike most strategies discussed in Section~\ref{sec:bg}, \name does not require model refactoring to work. In fact, PyTorch users can enable \name with few lines of code change to their existing training pipeline as shown in Figure~\ref{fig:ease-of-use}, allowing to train 10x larger models easily. For detailed tutorial, please see:  \url{https://www.deepspeed.ai/tutorials/zero-offload/}.

%Data scientists can adapt their model with few line of code change, and immediately take advantage of the memory savings that ZeRO offload offers.

\textbf{Contributions} Our contributions are as follows:
\begin{itemize}[leftmargin=*]
    \item A unique optimal offload strategy for heterogeneous large model training on GPU + CPU system that enables 10x larger model on a single GPU without sacrificing efficiency (Sec.~\ref{s:optimal_strategy}, Sec.~\ref{sec:single_schedule}).
    \item Highly scalable multi-GPU design through i) a symbiotic combination of offload strategy with ZeRO powered data parallelism (Sec.~\ref{s:multi-gpu}), allowing \name to achieve near linear scalability, and ii) seamless integration with model-parallel training \cite{Megatron-LM}, enabling even larger models than using \name or model parallelism alone (Sec.~\ref{s:multi-gpu}). 
    \item Optimized CPU execution through: i) High-performance CPU Adam optimizer design and implementation offering over 6x speedup over state-of-the-art Adam implementations (Sec.~\ref{subsec:cpu-adam}), and ii) One-step delayed parameter update to overlap CPU parameter update with GPU forward and backward pass (Sec.~\ref{subsec:delayed-parameter-update}). %With these two optimizations \name can remain highly efficient even at small batch sizes.
    \item Open-source implementation of \name in PyTorch. 
    \item Extensive evaluation demonstrating i) \emph{Model Scale}: 10x increase in model size with up to 13B on a single GPU and 4x increase in model size over model parallelism with up to 70 B parameters on a DGX-2 node. ii) \emph{Efficiency}: Over 40 TFlops for a 10B parameters on a single NVIDIA V100, compared to 30 TFLOPS on the same GPU with 1.2B parameters, the largest model that can be trained without any CPU offloading. iii) \emph{Scalability}: Near-perfect linear scalability for a 10B parameter model on up to 128 GPUs. iv) CPU overhead reduction with our ADAM implementation with 6x speedup over PyTorch optimizer and up to 1.5X improvement in end-to-end throughput with delayed parameter update optimizations (Sec.~\ref{s:evaluation}).
\end{itemize} 
\section{Background and Related Work}
\label{sec:bg}
%\subsection{Memory Consumption in Large Model Training} \label{s:model_states}
\textbf{Memory consumption in large model training.} The full spectrum of memory consumption during DNN model training can be classified into two parts: i) model states and ii) residual states \cite{zero}. Model states include parameters, gradients, and  optimizer states (such as momentum and variances in Adam~\cite{adam}); Residual states include activations, temporary buffers, and unusable fragmented memory. 

%\begin{itemize}
%    \item Model states, which includes parameters, gradients, and  optimizer states (such as momentum and variances in Adam~\cite{adam_iclr05}).
    
%    \item Residual states which includes activations, temporary buffers, and unusable fragmented memory. 
%\end{itemize}

%For state-of-art large model training, 
Model states are the primary source of memory bottleneck in large model training. We consider the memory consumption due to model states for large transformer models such as Megatron-LM (8 billion) \cite{Megatron-LM}, T5 (11 billion) \cite{raffel2020exploring}, and Turing-NLG \cite{turing-nlg} (17.2 billion). They are trained with float-16 mixed precision training \cite{mixed_percision_training} and Adam optimizer \cite{adam}. 

Mixed precision training often keeps two copies of the parameters, one in float-16 (fp16) and the other in float-32 (fp32). The gradients are stored in fp16. In addition to the parameters and gradients, the Adam optimizer keeps track of the momentum and variance of the gradients. These optimizer states are stored in fp32. Therefore, training a model in mixed precision with the Adam optimizer requires at least 2 bytes of memory for each fp16 parameter and gradient, and 4 byte of memory for each fp32 parameter, and the moementum and variance of each gradient. In total, a model with $M$ parameters requires $16 \times M$ bytes of memory. Therefore, the model states for Megatron-LM, T5 and Turing-NLG require 128 GB, 176 GB and 284 GB, respectively, which are clearly beyond the memory capacity of even the current flagship NVIDIA A100 GPU with 80 GB of memory.

%%%%Remove the following to save space%%%%%%%%
%The residual states or more specifically the activation memory is a secondary memory bottleneck, with memory requirement growing sub-linearly with the model size. For the transformer based models described above, the model size is a function of hidden dimension (h) and the number of transformer layers (l). The number of model parameters depends quadratically on the hidden size and linearly on the number of layers. In big $O$ notation, the number of parameters is $O(h^2l)$. On the other hand, the activation memory grows linearly with both hidden size and number of layers, which is of the order $O(hl)$. Hence, in terms of memory consumption, the activation memory is sub-linear to the number of parameters.

Significant amount of work has been done in the recent years to enable large model training, which requires more memory than what is available on a single GPU to fit these model and residual states. These efforts can be classified broadly into two categories: i) scale-out training  and ii) scale-up training based approaches. We discuss them as follows.

%\subsection{Scale Out Large Model Training}
%\label{s:scale-out-training}
\textbf{Scale out large model training.} Scale-out training uses aggregate memory of multiple GPUs to satisfy the memory requirement for large model training. Two prominent examples of scale out training is model parallelism~\cite{NIPS2012_4687,Megatron-LM} and pipeline parallelism~\cite{huang2018gpipe, harlap2018pipedream}, both partitioning the model states and the residual states across multiple GPUs. Model parallelism~\cite{NIPS2012_4687,Megatron-LM} partitions the model vertically and distributes the model partitions to multiple GPU devices in order to train large models. Pipeline parallelism~\cite{huang2018gpipe, harlap2018pipedream} on the other hand parallelizes model training by partitioning the model horizontally across layers. Both of these approaches must change the user model to work, therefore can limit usability.

A recent work, ZeRO \cite{zero}, provides an alternative to model and pipeline parallelisms to train large models. ZeRO splits the training batch across multiple GPUs similar to data parallel training \cite{NIPS2012_4687, parallel_sgd, jmlr20_data_parallelism}, but unlike data parallel training which replicates all the model states on each GPU, ZeRO partitions them across all GPUs, and uses communication collectives to gather individual parameters as needed during the training. ZeRO does not require changes to the user model to work, making it more generic than model or pipeline parallel training. It also offers better compute efficiency and scalability.

Despite the ability of model and pipeline parallelisms, and ZeRO to train large models, they all require multiple GPUs such that the aggregate GPU memory can hold the model and residual states for training large models. In contrast, \name is designed to fit a larger model by offloading model states to CPU memory and can train a 10x larger model on a single GPU without sacrificing efficiency. When multiple GPUs are available, \name is designed to work together with ZeRO to offer excellent scalability, or in conjunction with model parallelism to fit even larger model sizes that is not possible with \name or model parallelism alone.

%\subsection{Scale Up Large Model Training}
%\label{s:scale-up-training}
\textbf{Scale up large model training.} Existing work scales up model size in a single GPU through three major approaches. The first approach trades computation for memory saving from activations (residual memory) by recomputing from checkpoints~\cite{chen2016training}. The second approach uses compression techniques such as using low or mixed precision~\cite{mixed_percision_training} for model training, saving on both model states and activations.  
The third approach uses an external memory such as the CPU memory as an extension of GPU memory to increase memory capacity during training~\cite{vDNN_micro16,SuperNeurons_ppopp18,AutoTM_asplos20,SwapAdvisor_asplos20,Capuchin_asplos20,Layrub, hpca21:ren}. %Each of these three approaches are orthogonal to each other and can be used in conjunction with each other to fit larger models than possible with each of these approaches individually. 

Our work, \name falls under the third approach. Unlike \name, the above efforts only offload data to CPU but not compute, and they use smaller models training. 
%increasing batch size for smaller models and are not designed for large model training. %Please see \textbf{heterogeneous DL training} in Sec.~\ref{s:intro} for more details. 
In contrast, a recent work called L2L~\cite{layer-to-layer} can enable multi-billion parameter training by managing memory usage in GPU layer by layer. In particular, L2L synchronously moves tensors needed in the upcoming layer into GPU memory for computation and keeps the rest of tensors into CPU memory for memory saving. In comparison on \name, it offers limited efficiency due to extra communication overhead, does not offer a way to scale out across devices, and requires model refactoring, making it difficult to use.

\textbf{ZeRO powered data parallel training.} 
\name works with ZeRO to scale DL training to multiple GPUs. 
%To make the relation clear to the reader, here, we offer a quick overview of ZeRO. As mentioned earlier, ZeRO splits the training batch across multiple GPUs similar to data parallel training \cite{10.1109/4434.708258}, but unlike data parallel training which replicates all the model states on each GPU, ZeRO partitions them across all GPUs, and uses communication collectives to gather parameters as they are needed during the training. 
ZeRO has three stages, ZeRO-1, ZeRO-2 and ZeRO-3 corresponding to the partitioning of the three different model states, optimizer states, gradients and parameters, respectively. ZeRO-1 partitions the optimizer states only, while ZeRO-2 partitions gradients in addition to optimizer states, and ZeRO-3 partitions all model states. \name works symbiotically with ZeRO-2, and therefore we discuss it further. 

In ZeRO-2, each GPU stores a replica of all the parameters, but only updates a mutually exclusive portion of it during the parameter update at the end of each training step. As each GPU only updates a portion of the parameters, they only store optimizer states and gradients required to make that update. After the update, each GPU sends its portion of the updated parameters to all the other GPUs using an \texttt{all-gather}  communication collective. ZeRO-2 computation and communication schedule is described below: 

During the forward pass, each GPU computes the loss with respect to a different mini-batch. During the backward propagation, as each gradient is computed, it is averaged using a \texttt{reduce} operator at the GPU/GPUs that owns the gradient or part of the gradient. After the backward pass, each GPU updates its portion of the parameters and optimizer states using the averaged gradients corresponding to that portion. After this, an \texttt{all-gather} is performed to receive the rest of the parameter update computed on other GPUs.

\section{Unique Optimal Offload Strategy}\label{s:optimal_strategy}
\vskip -0.5em

\name is designed to enable efficient large model training on a single or multiple GPUs by offloading some of the model states from GPU to CPU memory during training. As discussed in Sec.~\ref{sec:bg}, model states: parameters, gradients, and the optimizer states, are the primary source of memory bottleneck in large model training. By offloading some of these model states to CPU, \name can enable training of significantly larger models \footnote{\name only offloads model states. Offloading secondary sources of memory bottleneck such as activation memory is beyond scope of our offload strategy. Given that they are significantly smaller than model states, we ignore them for the purpose of our analysis. Furthermore, the first and second approaches described in Sec.~\ref{sec:bg} can be used in conjunction with \name to reduce activation memory }. However, identifying the optimal offloading strategy is non-trivial. There are numerous ways to offload model states to CPU memory, each with a different trade-off in terms of CPU computation, and GPU-CPU communication, both of which can limit the training efficiency. 

To identify the optimal offload strategy, \name models the DL training as data-flow graph and uses first principle analysis to efficiently partition this graph between CPU and GPU devices. \name partitions the graph in a way that is optimal in three key aspects: i) it requires orders-of-magnitude fewer computation on the CPU compared to GPU that prevents CPU from becoming a performance bottleneck, ii) it guarantees the minimization of communication volume between CPU and GPU memory, and iii) it provably maximizes the memory savings while achieving minimum communication volume. In fact, \name can achieve high efficiency during training that is comparable to non-offload training and it is \emph{unique optimal}, meaning no other solution can offer better memory savings without increasing the communication volume or increasing CPU computation.

In this section, we discuss the derivation of our unique optimal offload strategy. Our strategy is specifically designed for \emph{mixed precision training with Adam optimizer} which is the de facto training recipe for large model training.

\vskip -4em
\subsection{DL Training as a data-flow graph}
\vskip -0.5em

The DL training workload can be represented as a weighted directed graph of data and computation, as shown in the figure~\ref{fig:dl_dataflow}, where the circular nodes represents model states (parameter16, gradient16, parameter32, momentum32, variance32), and the rectangular nodes represents computation (forward, backward, param update). The edges in the graph represents the data flow between the nodes, and the weight of an edge is the total data volume in bytes that flows through it during any given training iteration. For a model with M parameters, the weight of the edges in this graph is either $2M$ where the source node produces fp16 model states, or $4M$ where the source node produces fp32 model states. 

An offload strategy between GPU and CPU can be represented using a two-way partitioning of this graph, such that computation nodes in a partition would be executed on the device that owns the partition, and the data nodes in the partition will be stored on device that owns the partition. The total data volume that must be communicated between the GPU and CPU is given by the weight of edges running across two partitions. 

There are numerous ways to partition this graph. In the following sections, we use first principles to simplify the data flow graph to reduce the number possible choices based on three different efficiency metric: i) CPU computation overhead, ii) communication overhead, and iii) memory savings.

\subsection{Limiting CPU computation}
\label{s:limit-cpu}
\vskip -0.5em

The CPU computation throughput is multiple orders of magnitude slower than the GPU computation throughput. Therefore, offloading large computation graph to CPU will severely limit training efficiency. As such, we must avoid offloading compute intensive components to the CPU.

The compute complexity of DL training per iteration is generally given by $O(MB)$, where $M$ is the model size and $B$ is the effective batch size. To avoid CPU computation form becoming a bottleneck, only those computations that have a compute complexity lower than $O(MB)$ should be offloaded to the CPU. This means that the forward propagation and backward propagation both of which have a compute complexity of $O(MB)$ must be done on the GPU, while remaining computations such as norm calculations, weight updates etc that have a complexity of $O(M)$ maybe offloaded to the CPU. 

Based on this simple observation we fuse the forward and backward nodes in our data flow graph into a single super-node (FWD-BWD) and assign it on the GPU.

\begin{comment}
Fortunately, notice that the offloading strategy described above limits the total computation on CPU to O(M) while the computation on the GPU is O(MB). As batch size can be in thousands, \name not only achieves minimum communication volume while maximizing memory savings, it also severely limits the computation offloaded to the CPU, allowing \name to achieve similar training throughput to GPU only training, while enabling model sizes that are up to 10x larger.
\end{comment}

\subsection{Minimizing Communication Volume}

\begin{figure}[!tbp]
  \centering
   \includegraphics[ width=1.025\columnwidth]{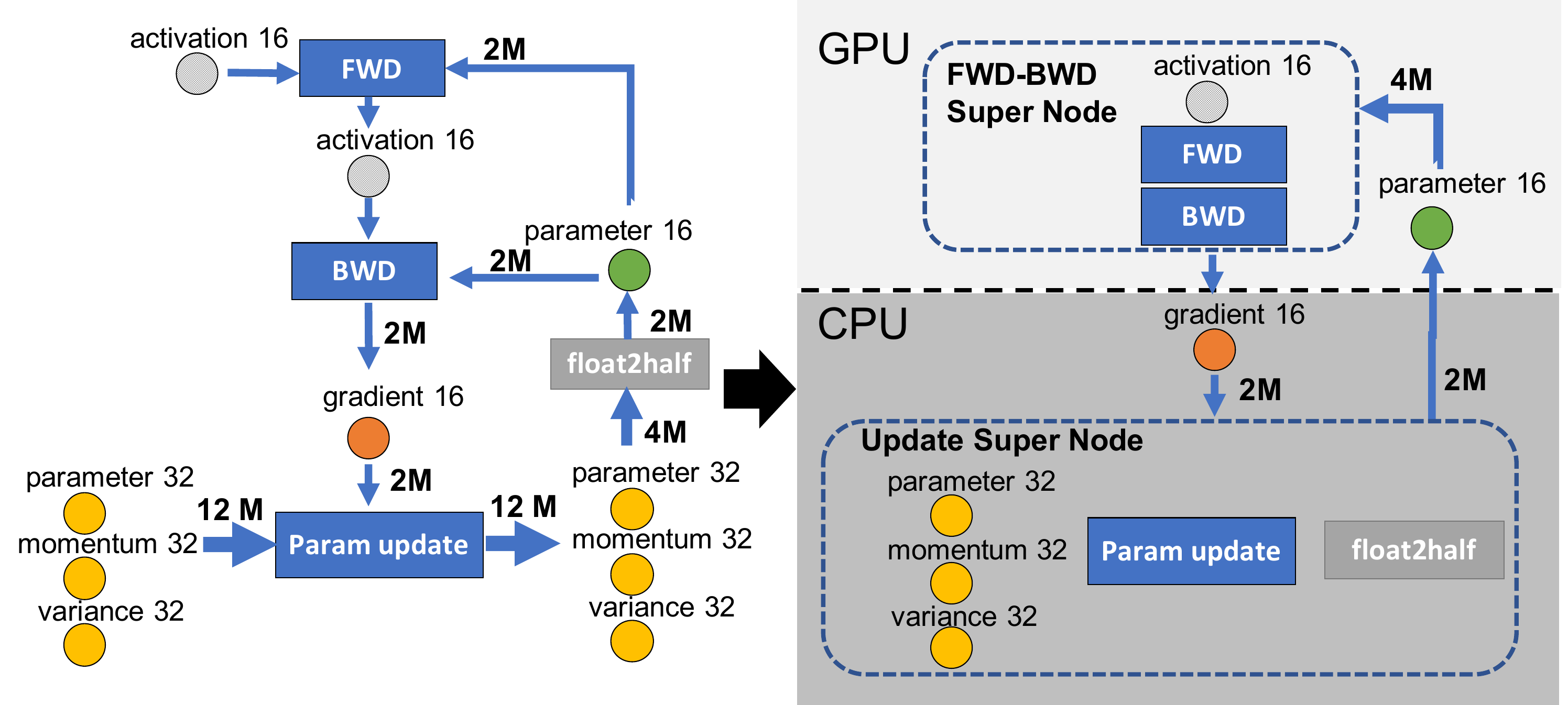}
   \vspace{-15pt}
  \caption{The dataflow of fully connected neural networks with $M$ parameters. We use activation checkpoint to reduce activation memory to avoid activation migration between CPU and GPU. }
   \vspace{-10pt}
\label{fig:dl_dataflow}
%\vspace{-10pt}
\end{figure}

The CPU memory bandwidth is at least an order of magnitude faster than the PCI-E bandwidth between CPU and GPU, while the GPU memory is another order of magnitude faster than even the CPU memory. Therefore, we must minimize the communication volume between CPU and GPU memory to prevent the PCI-E bandwidth from becoming a training performance bottleneck. To do so we must first identify the theoretical minimum communication volume for a model state offload strategy.

 The minimum communication volume for any model state offload strategy is given by $4M$ \footnote{Please note that it is possible to reduce the communication volume further by only offloading partial model states. For simplification, we assume that an offload of a model state implies that we offload the entire model state. Our analysis on the memory savings per communication volume, still holds even if we offload partial model states}. Note that after fusing the forward and backward into a single super-node as discussed in Sec.~\ref{s:limit-cpu}, each node in our data flow graph is part of a cycle. Therefore, any partitioning of this graph would require cutting at least two edges, each of which have a edge weight of at least $2M$, resulting in a total communication of at least $4M$.

If we choose to limit the communication volume to this bare minimum, we can greatly simplify our data-flow graph and reduce the number of partitioning strategies to a handful:  

\textbf{Creating fp32 super-node: } Notice that any partitioning strategy that does not co-locate the fp32 model states its producer and consumer nodes cannot achieve the minimum communication volume of $4M$. Such a partition must cut at least one edge with a weight of $4M$, and the other with at least $2M$, resulting in a communication volume of at least $6M$ . Therefore, to achieve the minimum communication volume, all offload strategies must co-locate fp32 model states with its producer and consumer operators, i.e., the fp32 model states (momentum 32, variance 32 and p32) must be co-located with the \emph{Param Update} and the \emph{float2half} computations.  

This constraint allows us to treat all the aforementioned fp32 data and compute nodes in the data flow graph as a single super-node that we refer to as \emph{Update Super}. We show this reduced data flow graph in figure~\ref{fig:dl_dataflow}, consisting of only four nodes: \emph{FWD-BWD Super} node, \emph{p16} data node, \emph{g16} data node, and \emph{Update Super} node.  

\textbf{p16 assignment: } To achieve the minimum communication volume, \emph{p16} must be co-located with \emph{FWD-BWD Super} because the edge weight between these two nodes is $4M$. Separating these two nodes, would increase the communication volume to $6M (4M + 2M)$. Since, we have already assigned node \emph{FWD-BWD Super} to GPU to limit computation on CPU, \emph{p16} must also be assigned to GPU.

\subsection{Maximizing Memory Savings}

After simplifying the data flow graph to minimize communication volume, only \emph{g16} and \emph{Update Super} remain to be assigned. Notice that at this point, all partitions will result in minimum communication volume, so we can prune the choices further to maximize the memory savings on the GPU. Table ~\ref{tab:memory-savings} shows the memory savings of all valid partitioning strategies that minimize the communication volume. The maximum memory savings of 8x can be achieved by offloading both \emph{g16} and \emph{Update Super} to CPU. 
\begin{table}[th!]
\scriptsize
\centering
\vskip -1em
\caption{Memory savings for offload strategies that minimizes communication volume compared to the baseline.}
\vskip -1em

\begin{tabular}{c|c|c|c|c|c}
     FWD-BWD&p16&g16&Update&Memory&Reduction  \\
     \hline
     gpu&gpu&gpu&gpu&16M&1x (baseline)\\
     gpu&gpu&cpu&gpu&14M&1.14x\\
     gpu&gpu&gpu&cpu&4M&4x\\
     gpu&gpu&cpu&cpu&4M&8x\\
\end{tabular}
\label{tab:memory-savings}
\vskip -2em

\end{table}

\subsection{A unique and optimal offload strategy}
\name allocates all the fp32 model states along with the fp16 gradients on the CPU memory, and it also computes the parameter updates on the CPU. The fp16 parameters are kept on the GPU and the forward and backward computations are also done on the GPU. 

We arrive at this offload strategy by simplifying our data flow graph and eliminating all other partitioning strategies as they do not limit CPU computation, minimize communication volume, or maximize memory savings. Therefore, \name is not only optimal in terms of the aforementioned metrics, it is also unique;  there can be no other strategy that can offer more memory savings than \name without increasing the compute complexity on the CPU or incur additional GPU-CPU communication volume.

\section{\name Schedule}
\label{sec:schedule}
\vspace{-5pt}

%Add a figure to show the workflow. If our design is not complicated, maybe we don't need overview.
In this section, we discuss the concrete computation and communication schedule for implementing \name on a single GPU system based on our offload strategy. We then show how we extend this schedule to work effectively on multi-GPU systems by combining our offload strategy with ZeRO data parallelism and model parallelism. 

\begin{comment}
ZeRO-Offload updates model parameters on the CPU, and therefore, requires efficient CPU optimizer to remain efficient. We find that state-of-art PyTorch implementation of CPU optimizers is sub-optimal. To address this, we discuss techniques for implementing an efficient CPU optimizer, and present the implementation of Adam optimizer on CPU as an example.  

Finally, we discuss the ZeRO-Offload API that empowers users to train large models requiring no more than few lines of code change, demonstrating the ease of use of ZeRO-Offload.
\end{comment}
\begin{figure}[!tbp]
  \centering
   \includegraphics[ width=\columnwidth]{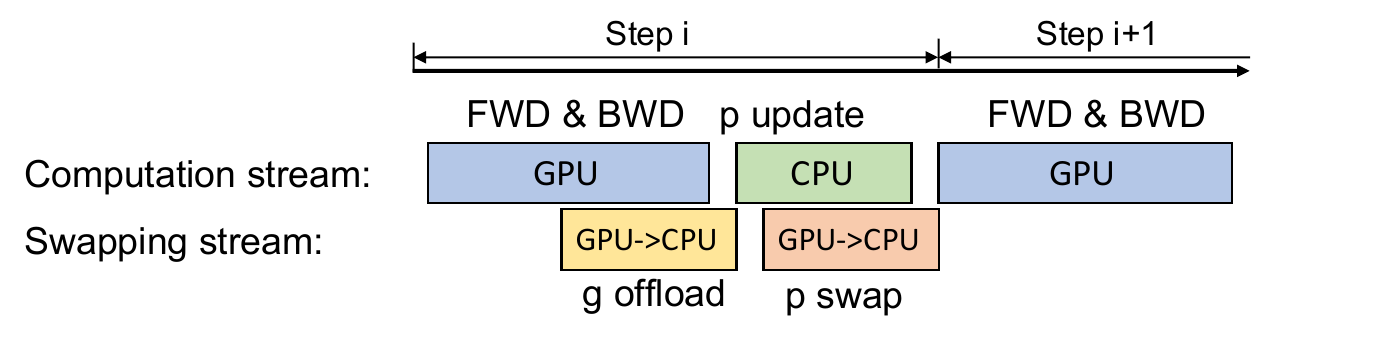}
    \vspace{-20pt}
  \caption{\name training process on a single GPU.}
   \vspace{-15pt}
\label{fig:zero-offload}
  % \vspace{-5pt}

\end{figure}
\subsection{Single GPU Schedule}
\label{sec:single_schedule}
\vspace{-5pt}

As discussed in Sec.~\ref{s:optimal_strategy}, \name partitions the data such that the fp16 parameters are stored in GPU while the fp16 gradients, and all the optimizer states such as fp32 momentum, variance and parameters are stored in CPU. 

During the training, we begin by computing the loss via the forward propagation. Since the fp16 parameters are already presented on the GPU, no CPU communication is required for this part of the computation. During the backward propagation on the loss, the gradient for different parameters are computed at different point in the backward schedule. \name can transfer these gradients for each parameter individually or in small groups to the CPU memory immediately after they are computed. Therefore, only a small amount of memory is required to temporarily hold the gradients on the GPU memory before they are transferred to CPU memory. Furthermore, each gradient transfer can be overlapped with the backpropagation on the remainder of the backward graph, allowing \name to hide a significant portion of the communication cost.

After the backward propagation, \name updates the fp32 parameters and the remaining optimizer states (such as momentum and variance) directly on the CPU, and copies the updated fp32 parameters from the CPU memory to the fp16 parameters on the GPU memory. Figure ~\ref{fig:zero-offload} shows the computation and communication in each step of \name diagrammatically,
and Figure~\ref{algo:pseudo-code} shows the concrete schedule as a pseudo-code.

\begin{figure}[!tbp]
  \centering
   \vspace{-5pt}

   \includegraphics[ width=\columnwidth]{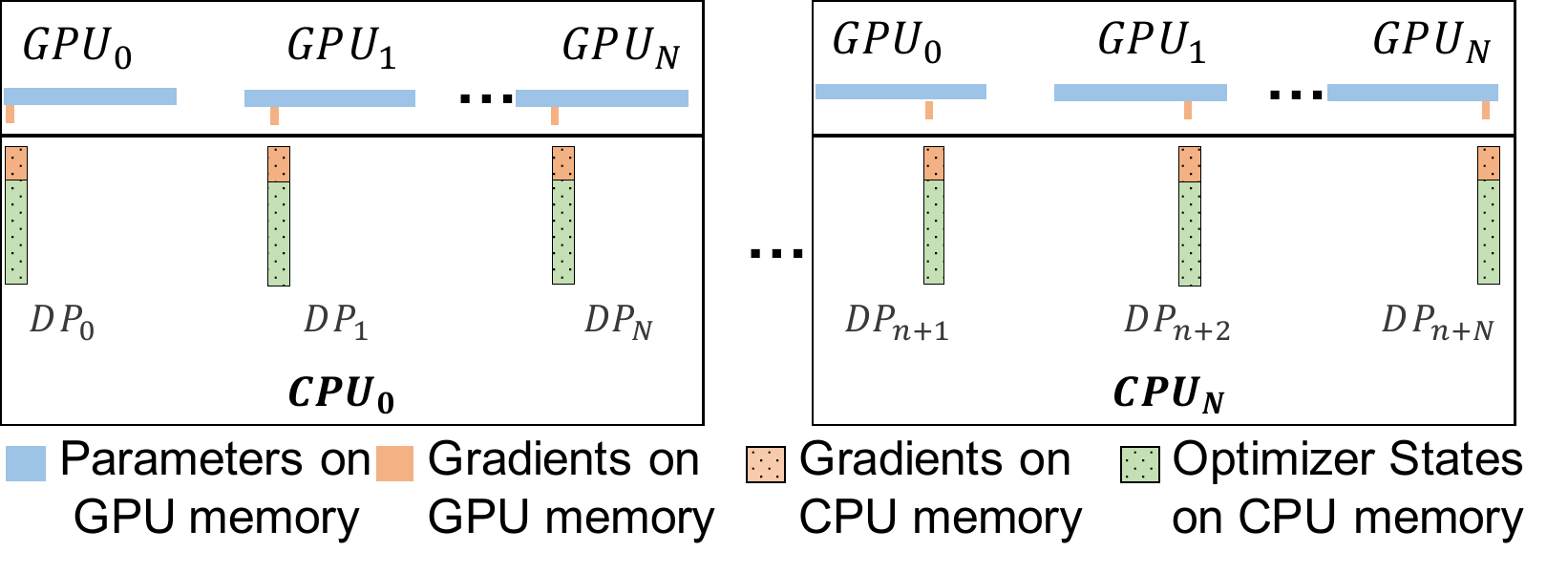}
    \vspace{-20pt}
  \caption{\name data placement with multiple GPUs}
  \vspace{-5pt}
\label{fig:data_placement}
 \vspace{-5pt}

\end{figure}
%\vspace{-5pt}
\subsection{Scaling to Multi-GPUs}
\label{s:multi-gpu}
%\vspace{-2pt}

\name in its entirety is a symbiotic integration of \name strategy described in Sec.~\ref{s:optimal_strategy} and ZeRO-powered data parallelism discussed in Sec.~\ref{sec:bg}, which allows \name  to scale to hundreds of GPUs efficiently. \name preserves the model state partitioning strategy of ZeRO Stage-2 (optimizer state and gradient partitioning), while offloading the partitioned gradients, optimizer states and the corresponding parameter updates to the CPU. 

The key benefit of doing this partitioning before offloading is that for systems with more than 1 GPU, each data parallel process is only responsible for updating a subset of the parameters. The aggregated communication volume from all the data parallel GPUs to the CPU remains constant, and CPU resources are used in parallel to jointly compute a single weight update. As a result the total CPU update time decreases with increased data parallelism, since the CPU compute resources increase linearly with the increase in the number of compute nodes. This allows \name to achieve very good scalability, as the overhead of communication across GPUs is offset by the reduction in the CPU optimizer step. 

\begin{figure}[t!]
  \centering
   \includegraphics[width=\columnwidth]{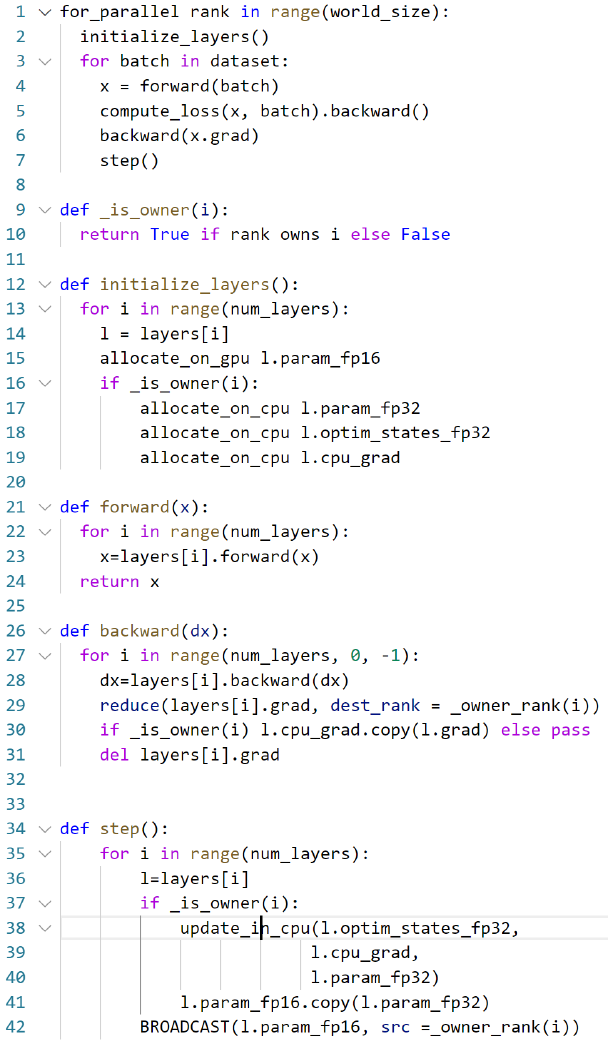}
    \vspace{-15pt}

  \caption{Code representing \name that combines unique optimal CPU offload strategy with ZeRO-powered data parallelism.}
\label{algo:pseudo-code}
 \vspace{-10pt}
\end{figure}

\name partitions gradients and optimizer states among different GPUs, and each GPU offloads the partition it owns to the CPU memory and keeps it there for the entire training. During the backward propagation, gradients are computed and averaged using reduce-scatter on the GPU, and each GPU only offloads the averaged gradients belonging to its partition to the CPU memory. Once the gradients are available on the CPU, optimizer state partitions are updated in parallel by each data parallel process directly on the CPU. After the update, parameter partitions are moved back to GPU followed by an all-gather operation on the GPU similar to ZeRO-2 to gather all the parameters. Figure~\ref{fig:data_placement} shows the data placement model parameters, gradients and optimizer states for \name and the details of the \name data parallel schedule is presented in Figure~\ref{algo:pseudo-code}. The all gather operation described above is shown as a sequence of broadcast operations in the Figure.

\begin{comment}

Notice that \name performs no redundant computation or communication. The CPU resources are used in parallel to jointly compute a single weight update, and the gradients are partitioned before offloading so only a single copy of gradients are sent to the CPU regardless of the number of GPUs. The major implication of this is that, the CPU-GPU communication bandwidth and CPU compute resources grows linearly with the number of GPUs, allowing \name to exhibit very good scalability. The efficiency and scalability is further improved by overlapping the communication and computation when possible to hide most of the offloading overhead. Please see Sec.~\ref{s:evaluation} for details.
\end{comment}

\textbf{Model Parallel training} \name can also work together with tensor-slicing based model parallelism (MP) frameworks such as Megatron-LM \cite{Megatron-LM}. It does so by offloading the gradients, optimizer states and the optimizer computation corresponding to each MP process allowing \name to train significantly larger models than possible than using model parallelism alone. %Please see Sec.~\ref{s:evaluation} for more details.  
Sec.~\ref{s:evaluation} provides more details.

\section{Optimized CPU Execution}
We speedup the CPU execution time for the parameter updates with two optimizations. 
First, we implement a fast CPU Adam optimizer using high performance computing techniques offering significant speedup over
state-of-art Pytorch implementation. Second, we develop a one-step delayed parameter update schedule that overlaps the
CPU parameter update computation with the forward and backward computation on the GPU, hiding the CPU execution time when enabled. 

\subsection{Implementing the CPU Optimizer}
\label{subsec:cpu-adam}

\begin{comment}
As discussed in Section~\ref{s:optimal_strategy}, we partition the computation in a way that has the minimum impact
on training throughput, by having the majority of computation processed on GPU, whereas offloading
the least blocking computing part to CPU. As the computation and communication dependency graph
of Figure~\ref{fig:dl_dataflow} illustrates, the optimizer (parameter update) requires the highest memory bandwidth. 
We observe that the PyTorch CPU optimizer implementations have a high latency resulting from poor CPU memory bandwidth utilization which limits the performance of ZeRO-Offload. Therefore, we implement our own CPU optimizers, which can significantly outperform PyTorch CPU optimizers. Here we discuss the techniques necessary for an efficient CPU optimizer, and discuss the implementation details using Adam as an example.
\end{comment}

\begin{comment}
Whereas, the optimizer updates each parameter individually and can be efficiently parallelized using vector instructions.
Therefore, we choose to do these operations on CPU using AVX instruction set, maintaining an equilibrium between the 
two optimization goals: maximize memory-saving on GPU and minimize performance overhead of the memory offload.
\end{comment}
We use three levels of parallelism for improving the performance of the CPU optimizer..
1) SIMD vector instruction \cite{SIMD-vector} for fully exploiting the hardware parallelism supported on CPU architectures. 
\begin{comment}
using this feature we can vectorize the independent parts of the computation, such as computing momentum, variance and updating each parameter. 
\end{comment}
2) Loop unrolling \cite{loop-unrolling}, an effective technique for increasing instruction level parallelism that is crucial for better memory bandwidth utilization. 
\begin{comment}by explicitly removing loop carried dependencies  dispatching the same operations multiple times to the vector units. This technique helps get better utilization of memory bandwidth, while pressuring the vector registers proportional to unrolling factor. We have tried different unrolling configurations and choose the one that gives us the best performance within the register limit.
\end{comment} 
3) OMP multithreading for effective utilization of multiple cores and threads on the CPU in parallel.
\begin{comment}
allowing to have several instances of the same code running in parallel on multiple CPU threads.
\end{comment}
Using these technique, we present a significantly faster implementation of Adam optimizer compared to state-of-art PyTorch implementation. 

\begin{comment}
used in large model training
Next, we present a fast implementation for ADAM optimizer as an example of how to alleviate the inefficiency of 
optimizer's runtime on CPU. We firstly go over the computations involved in the ADAM optimizer. Then, we describe the implementation 
detail by employing the vector instructions and OpenMP multithreading programming style.
\end{comment}
\paragraph{Mixed Precision Training with Adam}

%\begin{figure}[!tbp]
%  \centering
%   \includegraphics[ width=3.2in]{figures/cpu_adam.pdf}
%   %\vspace{-20pt}
%  \caption{An overview of CPU-ADAM implementation on CPU. 
%  We use three levels of parallelism, SIMD instructions, Loop unrolling,
%  and OMP multithreading for improving the efficiency of ADAM computation.}
%  %\vspace{-10pt} 
%\label{fig:cpu_adam}
%\end{figure}

ADAM is an optimization algorithm used for deep-learning training,
which takes the loss gradients together with their first and second momentums to update 
the parameters. Therefore, in addition to the model parameters, ADAM requires 
two more matrices of the same size ($M$) saved during the training. 
In the mixed precision training mode, there are two versions of the parameters stored in memory: 
one in FP16 (p16) used for computing the activations in the forward pass (on GPU), and
one master copy in FP32 (p32) which is updated by the optimizer (on CPU). The p16 is 
updated with the p32 through $float2half$ casting, at each training step. Moreover, the momentum
and variance of the gradients are saved in FP32 (on CPU), to prevent the precision loss for 
updating the parameters. Please refer to~\cite{adam} for further detail on ADAM's algorithm.
%The following equation show the computation of ADAM at each training step:
%\begin{equation}
%    \begin{array}
%        /m_t = \beta_1 m_{t-1} + (1-\beta_1)\nabla p_t \\
%        v_{t} = \beta_2 v_{t-1} + (1-\beta_2)\nabla p_t^2 \\
%        p_t = p_{t-1} - \alpha \times m_t/(1-\beta_1^t)/\sqrt{v_t/(1-\beta_2^t) + \epsilon}
%    \label{e:ADAM}
%    \end{array}
%\end{equation}

\begin{algorithm}\caption{CPU-ADAM Optimizer}\label{alg:cpuadam}
\vskip -1em
\caption{CPU-ADAM Optimizer}
\begin{flushleft}
         \scriptsize \textbf{Input:} $p32$, $g32$, $m32$, $v32$, $\beta_1$, $\beta_2$, $\alpha$ , $step$, $eps$ \\
         \textbf{Output:} $p16$, $p32$, $m32$, $v32$ \\
         \textbf{Parameter:} $tile\_width$, $simd\_width$, $unroll\_width$
\end{flushleft}
\vskip -1em

\begin{algorithmic}[1]
\vskip -0.1em

    \scriptsize \State $biascorrection1 \gets -\alpha/(1-\beta_1^{step})$
    \State $biascorrection2 \gets 1 / \sqrt{1 - \beta_2^{step}}$
    \State $simd\_count\gets sizeof(32)$ / $simd\_width$
   \State \textbf{unroll omp parallel}\For{i in 1 to ($simd\_count / unroll\_width$)}
                \State { ...
                \State $g_v$, $p_v$, $m_v$, $v_v$ = $g32[i], p32[i], m32[i], v32[i]$
                \State $m_v$ = FMA($g_v$, (1 - $\beta_1$), $\beta_1$*$m_v$)
                \State $v_v$ = FMA($g_v$*$g_v$, (1 - $\beta_2$), $\beta_2$*$v_v$)
                \State $g_v$ = FMA($\sqrt{v_v}$, $biascorrection2$, $eps$)
                \State $g_v$ = $m_m$ / $g_v$
                \State $p_v$ = FMA($g_v$, $biascorrection1$, $p_v$)
                \State $p32[i], m32[i], v32[i]$ = $p_v$, $m_v$, $v_v$
                \State ... }
                \State \textbf{IF} (i == tile\_width)  Copy\_to\_GPU(p16, p32)
    \EndFor
\end{algorithmic}
\end{algorithm}%\vskip -2em

\textbf{Optimized Implementation} Algorithm~\ref{alg:cpuadam} elaborates the ADAM's implementation detail using SIMD operations. As shown, the Adam function
receives the optimizer parameters such as $\beta_1$, $\beta_2$, and $\alpha$, and the gradient, momentum, variance and master copy of parameters (p32) as the input. We also use some parameters specific to the implementation, like the $simd\_width$ and $unroll\_width$.
The Adam optimizer sends back the updated variance, momentum, and parameter in both FP16 (to GPU) and FP32 (to CPU) .

We firstly read the data, including parameter, gradient, momentum and variance, into
the vector registers (line 7). Then, we use several fused multiply-add (\textit{FMA}) vector operations to preform the main
execution pipeline which is repeated by the unrolling width. Note that the rest of operations, such as multiply, division, and sqrt, 
also run in vector mode. For the best performance we use AVX512 simd instruction set and an $unroll\_width$ of 8 based on auto-tuning results. 
\begin{comment}
, We have tried different $unroll\_width$, 1, 2, 4, and 8 in order to achieve higher throughput and better utilize CPU memory bandwidth. 
Our numbers show the best performance when setting $unroll\_width$ to 8. 
Regarding the $simd\_width$, we have tried both Intel-x86 supported architectures, AVX2 and AVX512.
We have seen almost 10\%  performance improvement with AVX512, and we choose it for the rest of our experiments.
\end{comment}

In addition to the CPU-Adam optimizer, we implement the CPU-to-GPU 16FP parameter-copy in a tiled manner 
(line 15). We overlap the CPU and GPU execution by parallelizing the Adam computation and copying the 
parameters over to GPU. As we process Adam computation of the current tile of data on CPU, we write the parameters back 
to GPU for the previously processed tile. This way, we reduce the idle time of GPU to start the processing of the next training step.
%Moreover, we use the pinned memory for storing the FP32 parameter in order to exploit the memory-bandwidth between CPU and GPU

\subsection{One-Step Delayed Parameter Update}
\label{subsec:delayed-parameter-update}
\vskip -0.2em

\begin{figure*}
\centering
  % include first image
  \includegraphics[width=\linewidth]{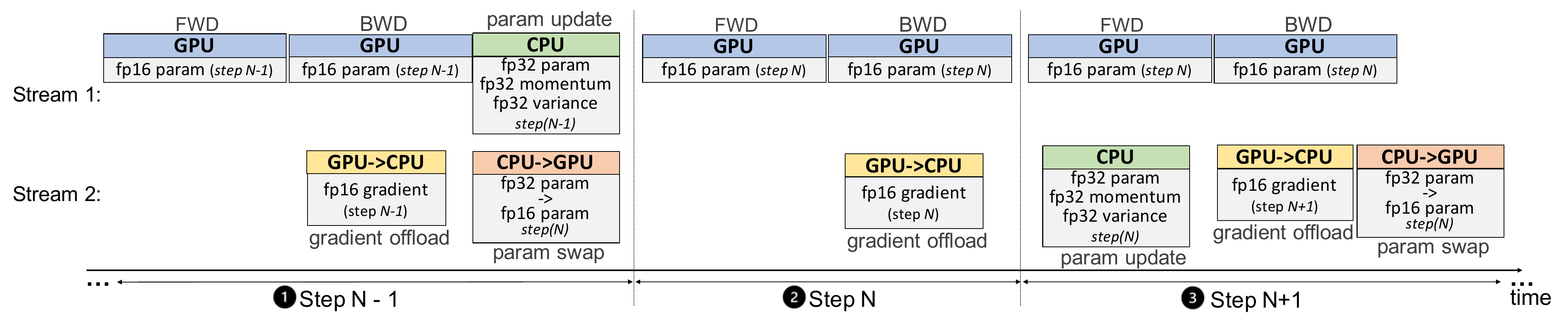}  
  \vspace{-20pt}
 \caption{Delayed parameter update during the training process.}
 \vspace{-5pt}
\label{fig:delay_param_update}
 \vspace{-5pt}
\end{figure*}

% \minjia{Let's call this section "Delayed Parameter Update"? I found it easier to cal it this way and to use DPU to represent the technique in the evaluation.}
Despite using a highly optimized CPU optimizer, the CPU computation overhead can become a bottleneck during training with very small batch sizes, when the GPU computation time is not much larger than CPU compute. For such limited cases, we develop one-step delayed parameter update (DPU) that overlaps CPU and GPU compute to hide the CPU computation overhead by delaying the parameter update by a single step. We verify that DPU does not impact the final accuracy of training in the evaluation.

\textbf{DPU training schedule} Figure~\ref{fig:delay_param_update} shows the workflow of \name training process with delayed parameter update. \ding{202} The first $N-1$ steps, are trained without DPU to avoid destabilizing the training during the early stages where gradients change rapidly.  
\begin{comment}
The training starts with update parameter timely. This is because at the beginning of the training process, the gradient has a rapidly descent (i.e., the gradient changes quickly). After first $N-1$ training steps finish, the CPU memory contains fp32 parameter, fp32 momentum and fp32 variance for used for the computation of step $N$ and the GPU memory contains updated fp16 parameters for step $N$. 
\end{comment}
\ding{203} On step $N$, we obtain the gradients from the GPU, but we skip the CPU optimizer step, and do not update the fp16 parameters on the GPU either. 
\begin{comment}
We skip parameter update on the We enable delayed parameter update when the gradient has a gentle descent (i.e., usually after first few training steps.) \name delays parameter update for only one step. For training step $N$, the forward computation and backward computation happens sequentially but do not execute optimizer computation. As described in figure~\ref{fig:delay_param_update}, we skip one optimizer computation once the parameter delay update enables. Meantime we offload gradients in step $N$ in CPU memory. 
\end{comment}
\ding{204} At step $N+1$, we compute the parameter updates on the CPU using gradients from step $N$, while computing the forward and backward pass on the GPU in parallel using parameters updated at step $N-1$. From this step onwards, 
the model at $(i+1)^{th}$ step will be trained using the parameters updated with gradients from $(i-1)^{th}$ step instead of parameters updated at $i^{th}$ step, overlapping CPU compute with GPU compute.

\begin{comment}
Figure~\ref{fig:delay_param_update} shows the training process after enabling delayed parameter update. For example in training step $N+1$, the computation for forward and optimizer happens simultaneously - \name uses fp16 parameter for forward computation on GPU, while \name uses fp32 momentum, fp32 variance and gradients stores on CPU memory to in-place update fp32 parameter on CPU. Backward computation uses fp16 parameter and gets updated fp16 gradient on GPU. \name offloads updated fp16 gradient from GPU memory to CPU memory. \name offloads gradient once the gradient computation finishes to save valuable GPU memory. Meanwhile \name updates the corresponding fp16 parameter based on updated fp32 parameters at same time. 
The following training steps keep update parameter asynchronously. 
\end{comment}
%follow the direction of the slope of the surface created by the objective function downhill until we reach a valley 

%\textbf{Select parameter update delay starting point. }

\textbf{Accuracy trade-off.} Since DPU changes the semantics of the training, it is reasonable to ask if there is a trade-off between model accuracy and training efficiency. To answer this question, we evaluated DPU on multiple training workloads and found that DPU does not hurt convergence if we introduce DPU after a few dozen iterations instead of introducing it at the beginning. Our evaluation result in Sec.~\ref{s:evaluation} shows that compared with training with \name only, training with delayed parameter update achieves same model training accuracy with higher training throughput. 
\begin{comment}
- %if we enable parameter update delay in early training stage, we could achieve higher training throughput but hurt model quality;
We find that if enabling delayed parameter update at the very beginning of training achieves high training throughput but also hurt model convergence speed and quality;
%if we enable parameter update delay in the middle or in the end of training process, we can not fully enjoy the high training throughput but get a model has the same accuracy as the model training without parameter update delay. 
if we enable delayed parameter update at a later stage, we may not reap the additional performance improvement from this optimization. 
%We enable parameter update delay after first 40 training steps. 
In practice, we find that enabling delayed parameter update after 40 training steps often provides significant throughput improvement without hurting model convergence.
\end{comment}

% \minjia{Broken section. The multi-GPU schedule section should probably go into Section~\ref{sec:schedule}.}

\section{Evaluation}
\label{s:evaluation}
\vskip -0.8em

This section seeks to answer the following questions, in comparison to the state-of-the-art:

\begin{enumerate}[label=(\roman*)]
    \item How does \name scale the trainable model size compared to existing multi-billion parameter training solutions on a single GPU/DGX-2 node?
    \item What is the training throughput of \name on single GPU/DGX-2 node?
    \item How does the throughput of \name scale on up to 128 GPUs?
    \item What is the impact of our CPU-Adam and delay parameter update (DPU) on improving throughput, and does DPU change model convergence?
\end{enumerate}

\subsection{Evaluation Methodology}
\begin{table}[tbp]
\footnotesize
\caption{Hardware overview of experimental system.}
\vskip -1em
\begin{tabular}{ll}
\hline
\multicolumn{2}{c}{\textbf{DGX-2 node}}         \\ \hline
GPU             & 16 NVIDIA Tesla V100 Tensor Core GPUs       \\
GPU Memory           & 32GB HBM2 on each GPU       \\
CPU              & 2 Intel Xeon Platinum 8168 Processors         \\
CPU Memory      & 1.5TB 2666MHz DDR4                            \\
CPU cache & L1, L2, and L3 are 32K, 1M, and 33M, respectively\\
PCIe      &     bidirectional 32 GBps PCIe
                  \\ \hline 
\end{tabular}
%\vspace{-12pt}
\label{tab:hardware}
\vskip -1.5em
\end{table}

% \textbf{Testbed. }To study the efficiency of \name, we first test \name on a single GPU. \name enables training 13-billion-parameter model on single GPU. We further perform tests on \name with one DGX-2 node with multiple GPUs and multiple DGX-2 nodes.
% Table~\ref{tab:hardware} shows the details of experimental platform. 

% \textbf{Model for evaluation.} We evaluate GPT-2~\cite{radford2019language} with \name. We vary the hidden dimension and the number of layers to obtain models with a different number of parameters. Table~\ref{tab:model} shows the configuration parameters used in our
% experiments with additional details.

% Another model we evaluate is Bert-large ~\cite{devlin2019bert}. The configuration parameter of bert-large-uncased-whole-word-masking model is: 24-layer, 1024-hidden, 16-heads, 336M parameters. And it fine-tuned on SQuAD. 

\textbf{Testbed.} For the evaluation of model scale and throughput, we conduct our experiments on a single DGX-2 node, whose details are shown in Table~\ref{tab:hardware}. For the evaluation of throughput scalability, we conduct experiments on 8 Nvidia DGX-2 nodes connected together with InfiniBand using a 648-port Mellanox MLNX-OS CS7500 switch.

\noindent
\textbf{Workloads.} For the performance evaluation, we focus on evaluating GPT-2~\cite{gpt2} like Transformer based models~\cite{transformer}. We vary the hidden dimension and the number of Transformer blocks to obtain models with a different number of parameters. Note that scaling the depth alone is often not sufficient because it would make training more difficult~\cite{open-aiscaling}. Table~\ref{tab:model} shows the configuration parameters used in our experiments. 
% \minjia{@Jie/@Samyam, How is the batch size selected? They do not appear to be the maximum batch size for a given size model.}
% \textcolor{jie}{jie: We test \name w/o MP with the maximum batch size can be trained on a single GPU (i.e., 1-13 billion model). We test \name w/ MP with a fixed batch size as 8. \minjia{Addressed.}}

For convergence analysis, such as the delayed parameter update, we use GPT-2~\cite{gpt2} and BERT~\cite{devlin2018bert}, both of which are commonly used as pre-trained language models and have demonstrated superior performance in many NLP tasks (e.g.,  natural language understanding and inference) than recurrent neural networks or convolutional neural networks.
% GPT-2 employs a left-to-right generative transformer based architecture, consisting of stacked decoders of Transformer blocks. We vary the hidden dimension and the number of Transformer blocks to obtain models with a different number of parameters. 
We use BERT-large, same as the one from \cite{devlin2018bert}, which has 24-layer, 1024-hidden, 16-heads, and 336M parameters. Similar to \cite{zero,Megatron-LM}, we fine-tune BERT on the Stanford Question Answering Dataset (SQuAD)~\cite{SQuDA-rank-list}, which is one of the most widely used reading comprehension benchmark~\cite{squda}.
Unless otherwise stated, we follow the same training procedure and hyperparameter settings as in \cite{devlin2018bert,gpt2}.

\noindent
\textbf{Baseline.} We compare the effectiveness of \name with state-of-arts multi-billion parameter training solutions:
\begin{itemize}
    \item PyTorch DDP: This is the existing PyTorch Transformer implementation using DistributedDataParallel~\cite{pytorch-ddp}.
    % \minjia{@Samyam, do we have a code link to the unmodified PyTorch Transformer implementation?}
    \item Megatron~\cite{Megatron-LM}: One of the current state-of-the-art multi-billion parameter model training solutions, which employs model parallelism to train up to 8.3B parameter models using 512 GPUs.  
    \item L2L~\cite{layer-to-layer}: L2L enables training of deep Transformer networks by keeping one Transformer block at a time in GPU memory and only moves tensors in the upcoming Transformer block into GPU memory when needed. 
    % leaving the rest of the model in CPU memory. L2L needs to move the upcoming layer into GPU memory for computation synchronously. 
    % We use the implementation of L2L in ~\cite{layer-to-layer-PyTorch} for evaluation, which requires user model changes. We use implementation of GPT-2 in~\cite{gpt_self_imple} for L2L evaluation.
    \item ZeRO~\cite{zero}: ZeRO extends data parallelism by eliminating memory redundancies across multiple GPUs, allowing to train models up to 170B parameters with high training throughput using 25 DGX-2 nodes. We refer to the open-sourced implementation of ZeRO as ZeRO-2. ZeRO-2 achieves the SOTA results for large model training and is a strong baseline.
\end{itemize}

\begin{table}[]
\caption{Model configuration in evaluation.}
\vskip -0.7em
\scriptsize
\begin{tabular}{c|c|c|c|c}
\hline
 \# params& \begin{tabular}[c]{@{}c@{}}batch size\\ per GPU\end{tabular} &\begin{tabular}[c]{@{}c@{}}MP setting\\ in \name\end{tabular}  & \# layer & hidden size  \\ \hline \hline
1, 2 billion  &32 &1     &  20, 40        &  2048    \\ \hline
%2 billion &32   &1    &  40        &   2048 \\ \hline
4 billion &32  &1     &  64        & 2304 \\ \hline
6, 8 billion &16  &1    &  53, 72       &   3072 \\ \hline
%8 billion  &16  &1    & 72         &  3072   \\ \hline
10,11 billion &10,8  &1    &  50,55        &   4096    \\ \hline
12, 13 billion  &4 &1    &  60, 65        &   4096 \\ \hline
%13 billion  &4  &1   &  65         &  4096 \\ \hline
15 billion  & 8  &2  & 78         &4096 \\ \hline
20,40,60 billion  &  8 &2  &  25,50,75        & 8192     \\ \hline
70 billion  & 8 &8   &  69        &9216 \\ \hline
\end{tabular}
\label{tab:model}
\vskip -2em
\end{table}

\subsection{Experimental Results}
% In the evaluation, we first compare \name with PyTorch and L2L on a single GPU testing platform. We also compare \name with Megaton and ZeRO-2 on a DGX-2 node with 16 V100 GPUs. We deploy \name on multiple DGX-2 nodes for scalability study. We enable activation checkpointing in \name, ZeRO-2 L2L, and Megatron-LM to save GPU memory. Same as \name, we also use mixed-precision training in ZeRO-2, L2L, and Megatron-LM. To comprehensively understand \name, we further evaluate CPU-ADAM efficiency in the end of this Section. 

\subsubsection{Model scale} 

% \name enables training 13-billion-parameter model on single GPU. We further perform tests on \name with one DGX-2 node with multiple GPUs and multiple DGX-2 nodes.

% evaluate the model trained on one GPU,  with \name and baseline solutions.
% The result shows in figure~\ref{fig:model_size}. \name enables 13B model training on a single GPU, which is 10x larger than using PyTorch for training. Megatron-LM and ZeRO-2 achieve the same model size as PyTorch, since both Megatron-LM and ZeRO-2 require multiple GPUs to partition the model memory saving.
% Moving unused layers to CPU memory empowers L2L to train larger models on a single GPU. 
% L2L enables 17 billion parameters modeling training on a single.

% We also compare the model that can be trained on one DGX-2 node with 16 GPUs. PyTorch with data parallelism does not increase the model size that can be trained on one DGX-2 node. ZeRO-2 partitions gradients and optimizer states among 16 GPUs, which enables training of the 8-billion-parameter model. 
% Megatron-LM leverages model parallelism, which can achieve 15-billion-parameter model training. L2L does not work with model parallelism. The increasing number of GPUs is not helpful for larger model training.  Combining with model parallelism, \name enables training of the 70-billion-parameter model on one DGX 2 node, which is 50x, 7.8x, 4.5x, and 4.2x larger than that trainable with PyTorch using data parallelism, ZeRO-2, Megatron-LM, and L2L, respectively. 

As an important step toward democratizing large model training, in this part, we first test the largest trainable models on a single GPU as well as 16 GPUs in a single DGX-2 node. 

\begin{figure*}
\begin{minipage}[t]{0.32\textwidth}
\centering
   \includegraphics[width=\columnwidth]{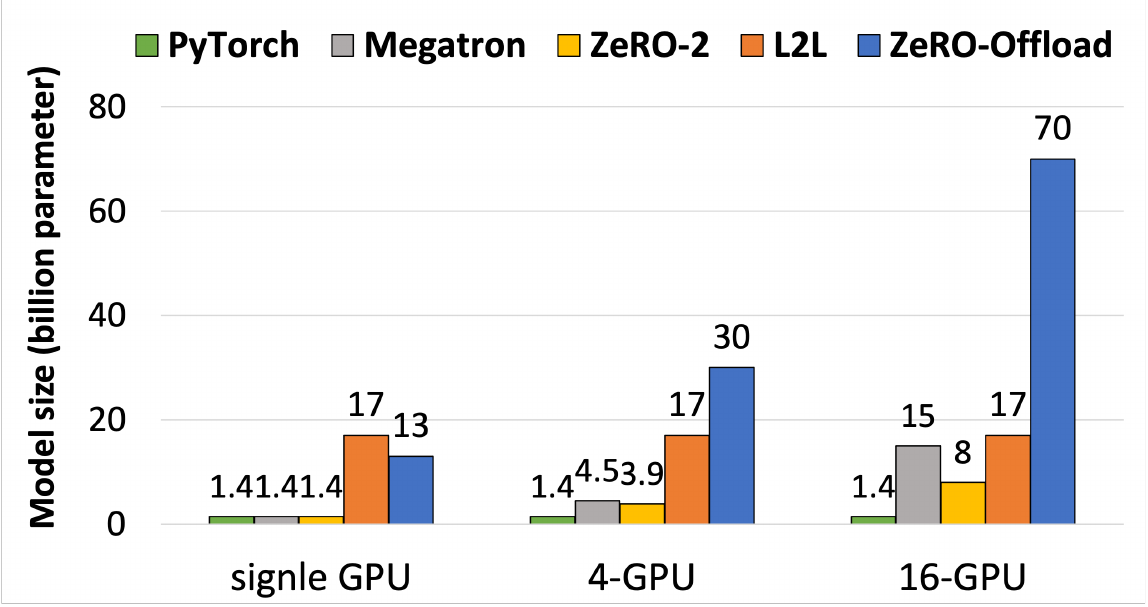}
   \vspace{-15pt}
  \caption{The size of the biggest model that can be trained on single GPU, 4 and 16 GPUs (one DGX-2 node).
  }
\label{fig:model_size}
\vspace{-5pt}
\end{minipage}
~
\begin{minipage}[t]{0.32\textwidth}
\centering
    \includegraphics[width=\columnwidth]{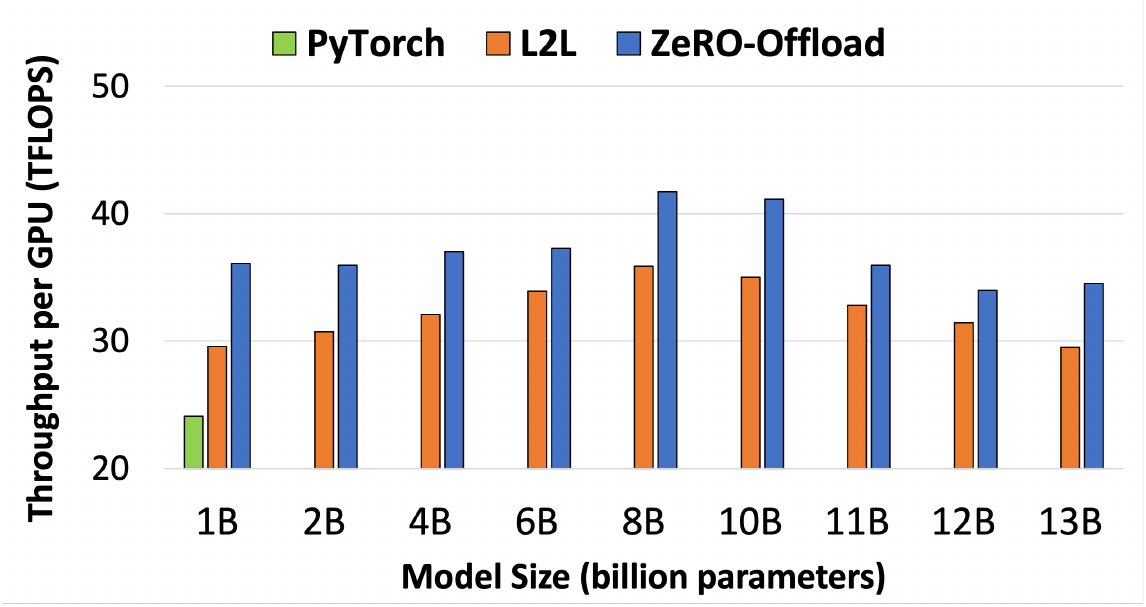}
    \vspace{-15pt}
    \caption{The training throughput with PyTorch, L2L, and \name on a single GPU with a batch size of 512. }
    %\minjia{TODO: In the title, "ZeRO-offload" -> "ZeRO-Offload".}
    %\textcolor{jie}{done}
    %\minjia{TODO: Change Y-axis to "Throughput per GPU (TFLOPS). Change X-axis to "Model size (billion parameters)}
    \label{fig:performance_one_gpu}
    \vspace{-5pt}
\end{minipage}
~
\begin{minipage}[t]{0.32\textwidth}
\centering
   \includegraphics[width=\columnwidth]{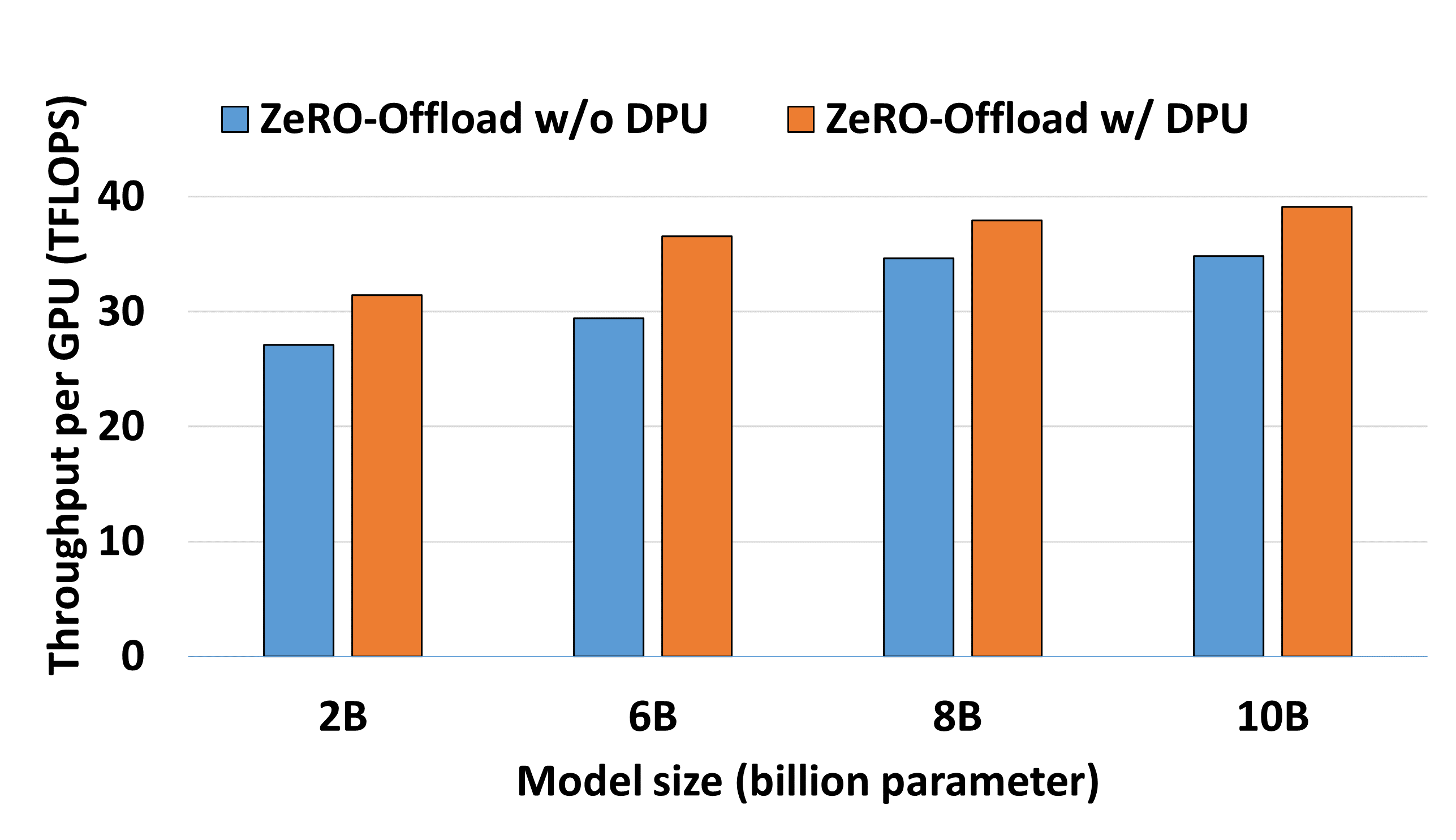}
   \vspace{-15pt}
  \caption{The training throughput is compared for w/o DPU and w/ DPU to GPT-2. Batch size is set to 8.}
  \label{fig:GPT2_2_6_8_10_B-delay-param-update}
  \vspace{-5pt}
\end{minipage}

\end{figure*}

\paragraph{Single GPU.} The largest model can be trained using PyTorch DDP on a single GPU with 32GB memory is 1.4B, before running out of memory, as shown in ~\ref{fig:model_size}.
Both Megatron and ZeRO-2 do not increase the trainable model size on a single GPU in comparison to PyTorch, because they both utilize the aggregated GPU memory to fit larger models. In contrast, \name enables 13B model training on a single GPU, which is more than 9X larger than using PyTorch, Megatron, and ZeRO-2. This is mainly because of \name's strategy for maximizing the memory savings on GPU by offloading expensive states such as optimizer states and the majority of gradients to CPU memory. On the other hand, L2L is able to train even larger models (e.g., 17B) on a single GPU by frequently moving weights from unused layers to CPU memory. However, the largest model size does not increase when training L2L with multiple GPUs, which is discussed next.
% L2L enables 17 billion parameters modeling training on a single.

% \minjia{The L2L paper says it can fit models up to 50 Billion parameters on a machine with a single 16GB V100 and 512GB CPU memory and without requiring any model partitioning. Why does Figure 7 show that L2L can only train model up to 17B with 32GB GPU memory?}
% \textcolor{jie}{jie: The maximum model size can be trained with L2L depends on the largest layer in that model. Compared with L2L paper, we are testing wider models. }
% \minjia{Addressed. This is subtle, L2L should be able to scale as long as there are sufficient CPU memory. However, in practice, it is not effective to scale just the model depth.}

\paragraph{Multi-GPU in single DGX-2.} We further perform model scale tests with 4 and 16 GPUs in a single DGX-2 node, respectively. As shown, the maximum trainable model size stays the same for both PyTorch and L2L, because
both of them do not handle memory redundancies in data parallelism. As a result, their scalability is bounded by the model scale on a single GPU. Both Megatron and ZeRO-2 support large model training with more GPUs, but they cannot scale efficiently beyond 15B parameters, even with 16 GPUs. Megatron supports larger models than ZeRO-2, because ZeRO-2 still incurs memory redundancies on model weights.
On the other hand, \name easily enables training of up to 70B parameter models by partitioning and offloading optimizer states and gradients to CPU memory combined with model parallelism. Overall, \name increases the model scale on a single DGX-2 node by 50X, 4.5X, 7.8X, and 4.2X than using PyTorch, Megatron, ZeRO-2, and L2L, respectively.  

% because the aggregated GPU memory (512GB) is still far smaller than the available CPU plus GPU memory (512GB + 1.5TB). 

% is able to train 15B models with its model parallelism scheme enabled. ZeRO-2 is only able to scale up to 8B parameter models, by partitioning gradients and optimizer states across 16 GPUs. With model parallelism, \name enables training of up to 70B parameter models on one DGX 2 node, which is 

% evaluate the model trained on one GPU,  with \name and baseline solutions.
% The result shows in figure~\ref{fig:model_size}.  Megatron-LM and ZeRO-2 achieve the same model size as PyTorch, since both Megatron-LM and ZeRO-2 require multiple GPUs to partition the model memory saving.
% 

% We also compare the model that can be trained on one DGX-2 node with 16 GPUs. PyTorch with data parallelism does not increase the model size that can be trained on one DGX-2 node. 
%  L2L does not work with model parallelism. The increasing number of GPUs is not helpful for larger model training.  
%%%%%%%%%%%%%%%%%%%%%%%%%%%%%%%%%%%%

\subsubsection{Training throughput} 

% \minjia{I assume these results are collected with 1-step delay disabled. Given that delayed parameter update is part of the ZeRO-Offload design, not sure if we should enable it when measuring the throughput. Current solution is to explicitly mention that these throughput results are collected with DPU, and DPU is separately evaluated in a later section.}

\paragraph{Single GPU.} Next, we compare the training throughput of \name and L2L, for models with billion-scale parameters, on a single GPU. We do not include Megatron and ZeRO-2 in this comparison, because both of them cannot train models bigger than 1.4B parameters due to OOM. 
% To have a fair comparison, 
We evaluate \name and L2L with the same training batch size (e.g., 512) and same micro-batch sizes (shown in table~\ref{tab:model}), with gradient accumulation enabled. We also disable delayed parameter update in this experiment so that the comparison is only from the system efficiency perspective. We evaluate the performance improvement and its impact on the convergence of delayed parameter update in Section~\ref{subsubsec:cpu-overhead}.

Figure~\ref{fig:performance_one_gpu} shows that \name outperforms L2L by 14\% on average (up to 22\%) in throughput (TFLOPS). The performance benefit of \name comes from the following two aspects. First, \name has a lower communication cost between CPU and GPU than L2L. For a model with $M$ parameters, L2L requires $28M$ data communication volume between GPU and CPU, which is a sum of the weights, gradients, and optimizer states of each layer of the model. As analyzed in Sec.~\ref{sec:single_schedule}, the communication volume between CPU and GPU memory in \name is $4M$, which is 7x smaller than L2L. The reduced communication volume significantly mitigates the bottleneck from CPU-GPU communication. 
Second, compared with L2L, the parameter update of \name happens on CPU instead of GPU, but our optimized CPU-Adam implementation achieves a quite comparable parameter update performance than the PyTorch Adam implementation on GPU (evaluated in Sec.~\ref{subsubsec:cpu-overhead}). Therefore, although the optimizer update on GPU in L2L is slightly faster than the optimizer update on CPU in \name, the communication overhead introduced by L2L leads to an overall slower throughput than \name. 

% \minjia{Need a revision once we have the PyTorch GPU Adam on L2L.}

% \emph{Original: Second, compared with \name, L2L, the optimizer step calculation on GPU is much faster than on CPU. With gradient accumulate update, \name significantly reduces the overhead of updating parameter per batch. While L2L micro-batch looping reduces limited overhead in updating parameters, it is not helpful to reduce communication overhead between GPU and CPU memory. }

% \minjia{@Jie/@Samyam/@Reza, the second point is not very clear. Are you saying that the CPU-Adam and gradient accumulation makes the optimizer update on CPU faster than GPU-Adam plus the L2L CPU-GPU communication time? }

%For the models which is larger than 10 billion parameter model, L2L and \name has comparable performance, this is because

\begin{comment}
\begin{figure}[!tbp]
  \centering
    \includegraphics[ width=\columnwidth]{figures/performance_one_gpu_new.pdf}
    \caption{Training throughput with PyTorch, L2L, and \name on a single GPU.}
    %\minjia{TODO: In the title, "ZeRO-offload" -> "ZeRO-Offload".}
    %\textcolor{jie}{done}
    %\minjia{TODO: Change Y-axis to "Throughput per GPU (TFLOPS). Change X-axis to "Model size (billion parameters)}
    \label{fig:performance_one_gpu}
\end{figure}
\end{comment}
\noindent
\textbf{Multi-GPU in single DGX-2. }
Next, we compare the training throughput of PyTorch, ZeRO-2, Megatron, \name without model parallelism (w/o MP), and \name with model parallelism (w/ MP) in one DGX-2 node. When using MP, we use a MP degree that gives the best performance for both baseline and \name. We use a total batch size of 512 for all the experiments using a combination of micro-batch per GPU and gradient accumulation. To get the best performance for each configuration, we use the largest micro batch that it can support without OOM.    
% We refer \name with inner node model parallel as \name + Megatron-LM. 
We exclude L2L~\cite{layer-to-layer-PyTorch} in this test because its implementation 
does not support multi-GPU training. 

\begin{figure}[!tbp]
  \centering %\includegraphics[width=\columnwidth]{figures/performance_one_node.pdf}
  \includegraphics[width=\columnwidth]{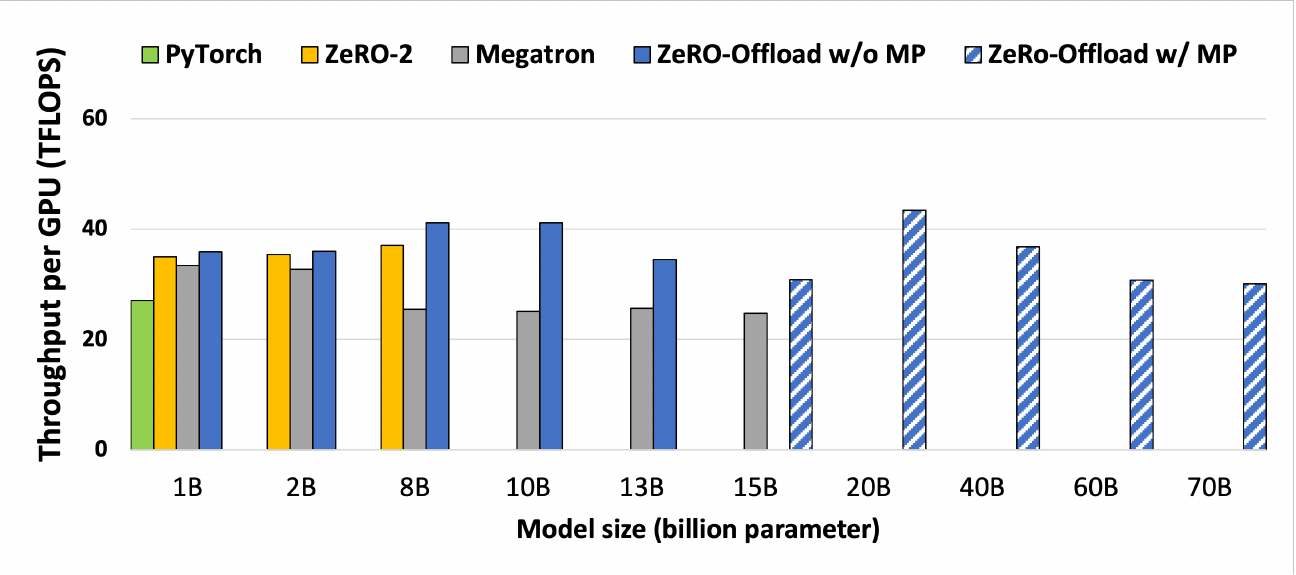}
   \vspace{-20pt}
  \caption{Training throughput with PyTorch, ZeRO-2, Megatron-LM, \name without model parallelism and \name with model parallelism.}
  \vspace{-10pt}
  %\minjia{@Shuangyan/@Jie, could you please update the throughput at 40B model size, based on @Reza's updated result?}
  %\minjia{TODO: Change "ZeRO-offload" to "ZeRO-Offload w/o MP". Change "ZeRO-offload + Megatron" to "ZeRO-Offload w/ MP.}\textcolor{jie}{done}
  %\minjia{TODO: Change "pytorch" to "PyTorch".}\textcolor{jie}{done}
  %\vspace{-10pt} 
\label{fig:performance_one_node}
  \vspace{-5pt}

\end{figure}

Figure~\ref{fig:performance_one_node} shows the throughput per GPU results when training on multiple GPUs. We make the following observations:
\begin{itemize}
    \item For 1B to 15B models, \name achieves the highest throughput and has up to 1.33X, 1.11X, 1.64X higher speeds than PyTorch, ZeRO-2, and Megatron, respectively. By offloading all the optimizer states to CPU with low overhead, \name can train with larger micro-batch sizes giving higher throughput.
    %\minjia{@Shuangyan, could you please add the relative speedups between \name and the other baselines?}
    \item ZeRO-2 runs out of memory once the model size is beyond 8B due to lack of enough aggregate GPU memory to store the model states on 16 GPUs. Instead, \name scales to 13B, without model parallelism because it offloads optimizer states and the majority of gradients to CPU memory.
    \item When combined with model parallelism, \name enables training up to 70B parameter models with more than 30 TFLOPS throughput per GPU. In contrast, Megatron supports only up to 15B parameter models before running out of memory, using just model parallelism.
    \item Compared \name with ZeRO-2 and Megatron, \name outperforms ZeRO-2 and Megatron in throughput for 1--8B and 1--13B parameter models, respectively. \name is faster than Megatron, because it eliminates frequent communication between different GPUs and can train with larger micro batch sizes. \name outperforms ZeRO-2 also due to larger micro batch sizes.
    %\item Training model with PyTorch Distributed Data Parallel has the lowest training throughput among all the %solutions, because it is restricted to have a smaller micro batch size (i.e., often has lower computation %efficiency) than other solutions due to the lack of GPU memory capacity.
\end{itemize}

%%%%%%%%%%%%%%%%%%%%%%%%%%%%%%%%%%
% \begin{figure*}[!tbp]
%   \centering
%   \includegraphics[width=1.75\columnwidth]{figures/gas_1.pdf}
%   \caption{\textcolor{jie}{Training throughput with PyTorch, ZeRO-2, Megatron-LM, ZeRO-Offload without model parallelism and ZeRO-Offload with model parallelism. effective batch size = 512}}
%   %\minjia{TODO: TFLOPS (e.g., this figure) and TFlops (e.g., Figure 10) are used interchalleably in all figures. Let's use TFLOPS in all figures.} \textcolor{jie}{done}
%   %\minjia{TODO: Change "ZeRO-offload" to "ZeRO-Offload" in the title.}\textcolor{jie}{done}
% \label{fig:performance_one_node_new}
% \end{figure*}

\vskip -1em
\subsubsection{Throughput Scalability}

\begin{figure}[!tbp]
  \centering
   \includegraphics[width=\columnwidth]{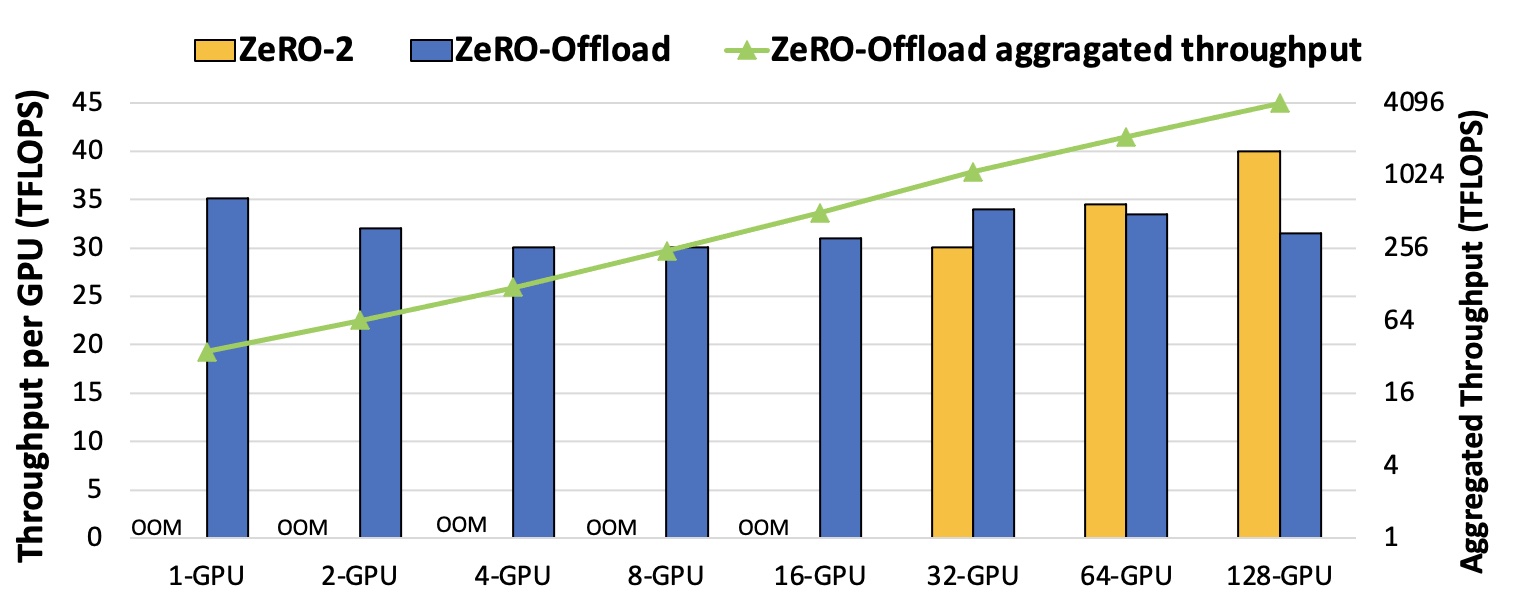}
   \vspace{-15pt}
  \caption{Comparison of training throughput between \name and ZeRO-2 using 1--128 GPUs for a 10B parameter GPT2.}
  %\minjia{TODO: TFLOPS (e.g., this figure) and TFlops (e.g., Figure 10) are used interchalleably in all figures. Let's use TFLOPS in all figures.} \textcolor{jie}{done}
  %\minjia{TODO: Change "ZeRO-offload" to "ZeRO-Offload" in the title.}\textcolor{jie}{done}
\label{fig:scalibility}
\vspace{-10pt}
\end{figure}

We compare the throughput scalability of ZeRO-2 and \name \footnote{We do not include comparison against Megatron because it consistently performs worse than \name, as shown in Figure~\ref{fig:performance_one_node}. Given the communication overhead added by model parallelism, scaling out Megatron training can not achieve higher throughput than \name even with linear scalability.} on up to 128 GPUs in Figure~\ref{fig:scalibility}
 %\textcolor{red}{XX}.
% \minjia{@Jie/@Samyam, Why do we only compare with ZeRO-2 but not Megatron in the scalability test?}
 and make the following key observations: First, \name achieves near perfect linear speedup in terms of aggregate throughput (green line) running at over 30 TFlops per GPU (blue bars). Second, from 1 to 16 GPUs, while ZeRO-2 runs out of memory, \name can effectively train the model, turning the model training from infeasible to feasible. Third, with 32 GPUs, \name slightly outperforms ZeRO-2 in throughput. The improvement comes from additional memory savings on GPU from \name, which allows training the model with larger batch sizes that lead to increased GPU computation efficiency. Fourth, with more GPUs (such as 64 and 128), ZeRO-2 starts to outperform \name, because both can now run similar batch sizes, achieving similar computation efficiency, whereas ZeRO-2 does not suffer from the additional overhead of CPU-GPU communication. 
% On one hand, though, ZeRO-2 does not have the overhead of moving data to CPU, while on the other hand, the optimizer step calculation on GPU is much faster than on CPU. 
In summary, \name complements ZeRO-2 and enables large model training from a single device to thousands of devices with good computation efficiency. 

\subsubsection{Optimized CPU execution}
\label{subsubsec:cpu-overhead}

\paragraph{A. CPU-Adam efficiency.}
In this part, we evaluate our Adam implementation against the PyTorch Adam on CPU. 
Table~\ref{t:adamperf} shows the optimizer execution time of the two implementations
for model parameters from 1 to 10 billion. Compared to PyTorch (PT-CPU), CPU-Adam reduces the execution time by over 5X for all the configurations and 6.4X for the case with 1B parameters. The CPU-Adam optimizer achieves high speedups by exploiting the instruction-level parallelism, thread-level parallelism, and the tile-based data copy scheme (as shown in line 15 of Algorithm~\ref{alg:cpuadam}). Meanwhile, although CPU-Adam has a slower speed than the PyTorch Adam implementation on GPU (PT-GPU), the performance gap is not very huge, and the CPU computation is not a bottleneck of the training throughout. 

\vspace{-15pt}

\noindent
\paragraph{B. One-step Delayed parameter update (DPU).} 

Figure~\ref{fig:GPT2_2_6_8_10_B-delay-param-update} shows the comparison of the training throughput of GPT-2 with and without DPU. As shown, with DPU enabled, the training achieves 1.12--1.59, updated times higher 
% \minjia{@Shuangyan, can you update these numbers based on the new results in Figure 11?}
throughput than without it, for a wide range of model sizes for a small micro batch size of 8. 
% \minjia{@Shuangyan, could you please add the speedups? }
This is expected because DPU allows the optimizer updates to overlap with the next forward computation such that the GPU does not have to be slowed down by the CPU computation and CPU-GPU communication. But, what about accuracy?

\begin{figure}[!htb]
%\hfill
\vskip -1em
\begin{minipage}{.22\textwidth}
   \includegraphics[width=1.5in, height=1.2in]{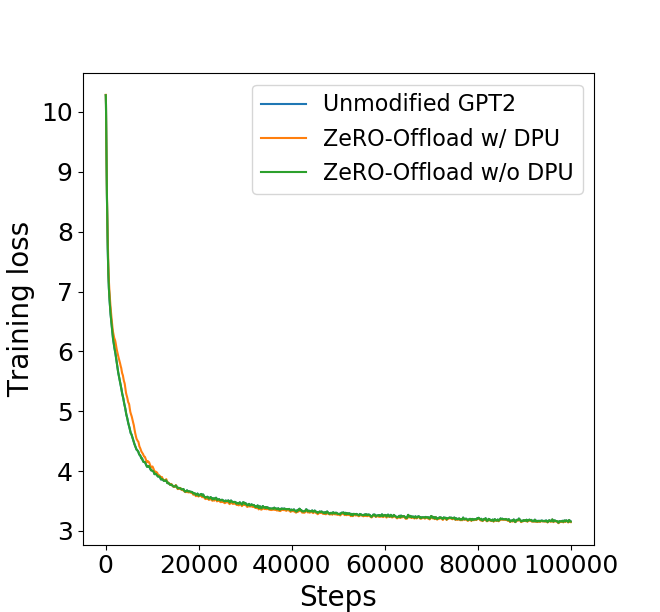}
  \caption{The training loss curve of unmodified GPT-2, \name w/o DPU and \name with DPU.}
  \vskip -1em

\label{fig:GPT-2_train_loss_curve}

%%\minjia{TODO: Increase the font of the x-axis, y-axis, and legend.}
%%\minjia{TODO: Just show two curves: "\name without DPU" and "\name with DPU". }
%%\minjia{TDOO: Remove the title "GPT-2".}
%%\minjia{It would be nice to have the convergence results from Tunji that show more iterations (basically until the model converges).}
\end{minipage}
~
\begin{minipage}{.22\textwidth}
%\vskip -1em
   \includegraphics[width=\columnwidth]{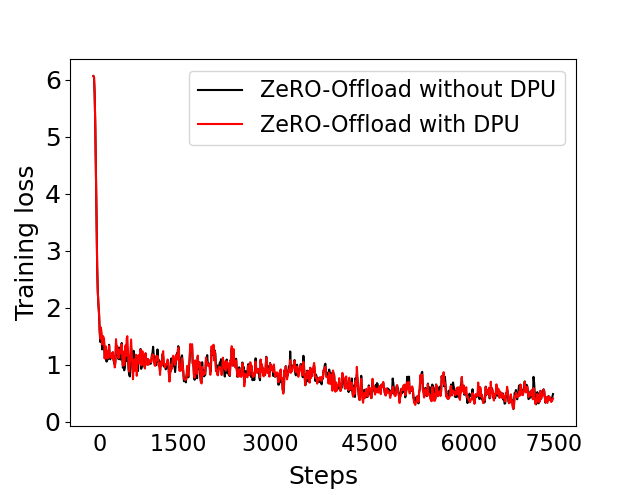}
    %   \includegraphics[width=\columnwidth]{figures/bert-large-loss-curve-smoothing-0.5.png}
%   \vspace{-15pt}
  \caption{The fine-tuning loss curve of BERT,  \name w/o DPU and \name with DPU.   }
  \vskip -1em

\label{fig:bert-large_loss_curve_smoothing-0.5}

% \minjia{TODO: Increase the font of the x-axis, y-axis, and legend.}
% \minjia{TODO: Just show two curves: "\name without DPU" and "\name with DPU". }
% \minjia{TDOO: Remove the title "bert-large...".}
% \minjia{TODO: Y-axis label changes to "Training loss".}
\end{minipage}
\vskip -1em

\end{figure}

\begin{comment}The delayed parameter update improves training throughput, but it may also have an impact on model accuracy because the forward computation uses weight parameters from the previous step, which introduces a source of staleness. Intuitively, the staleness has a bigger impact on model convergence in the beginning of training since the model is actively updating its parameters to learn new features. As the training proceeds, such staleness would have a diminishing impact as the learning rate decays and model weights become more stable.
\end{comment}
\textbf{Convergence impact} We study the convergence impact of DPU on both GPT-2 and BERT. Figure~\ref{fig:GPT-2_train_loss_curve} shows the pre-training loss curves over 100K training iterations using PyTorch (unmodified GPT-2), and Figure~\ref{fig:bert-large_loss_curve_smoothing-0.5} shows the loss curves of fine-tuning Bert-large model on SQuAD using \name without DPU, and \name with DPU. In both cases, DPU is enabled after 40 iterations allowing the training to stabilize in its early stage before introducing DPU. 

We observe that the training curves of the unmodified GPT-2 and \name w/o DPU are exactly overlapped, because \name w/o DPU performs only system optimizations and does not alter training dynamics. On the other hand, the training curve from \name with DPU converges slightly slower at the very beginning of the training (e.g., barely can be seen at 2K-5K iterations) and quickly catches up after 5K iterations. For the remaining of the training, the training loss matches the original training until the model converges. 

For Bert-Large fine-uning, we can see that although the training losses are not exactly the same, they converge in the same trend and are largely overlapped. Without changing any hyperparameters, \name + DPU achieves the same final F1 score (92.8) as the baseline.
\begin{table}[t!]
\caption{Adam latency (s) for PyTorch (PT) and CPU-Adam.}
\vspace{-5pt}
\label{t:adamperf}
\centering
\begin{tabular}{|c|c|c|c|c|}
\hline
\cellcolor[gray]{0.9} \scriptsize \textbf{\#Parameter}& \cellcolor[gray]{0.9} \scriptsize \textbf{CPU-Adam}& \scriptsize \cellcolor[gray]{0.9} \textbf{PT-CPU}& \scriptsize \cellcolor[gray]{0.9} \textbf{PT-GPU (L2L)}\\
\hline
\scriptsize 1 billion&\scriptsize 0.22&\scriptsize 1.39&\scriptsize 0.10\\
\hline
\cellcolor[gray]{0.9}\scriptsize 2 billion & \cellcolor[gray]{0.9}\scriptsize 0.51& \cellcolor[gray]{0.9}\scriptsize 2.75& \cellcolor[gray]{0.9}\scriptsize 0.26\\
\hline
\scriptsize 4 billion & \scriptsize 1.03& \scriptsize 5.71& \scriptsize 0.64\\
\hline
\cellcolor[gray]{0.9}\scriptsize 8 billion& \cellcolor[gray]{0.9}\scriptsize 2.41& \cellcolor[gray]{0.9}\scriptsize 11.93& \cellcolor[gray]{0.9}\scriptsize 0.87\\
\hline
\scriptsize 10 billion& \scriptsize 2.57& \scriptsize 14.76& \scriptsize 1.00\\
\hline
\end{tabular}
\vspace{-10pt}
% \minjia{@Reza, could you please change all the digits to have 1 significant digit after the decimal? E.g., 6.381 -> 6.4.\minjia{Done.}}
\end{table}
From these results on both GPT-2 pretraining, and Bert-Large fine-tuning, we empirically verify that DPU is an effective technique to improve the training throughput of \name without hurting model convergence and accuracy.The 1-step staleness introduced by DPU is well tolerated by the iterative training process once the model has passed the initial training phase.

\section{Conclusions}
We presented \name, a powerful GPU-CPU hybrid DL training technology with high compute efficiency and near linear throughput scalability, that can allows data scientists to train models with multi-billion parameter models even on a single GPU, without requiring any model refactoring. We open-sourced \name as part of the DeepSpeed library (\url{www.deepspeed.ai}) with the hope to democratize large model training, allowing data scientist everywhere to harness the potential of truly massive DL models. 

%-------------------------------------------------------------------------------

%-------------------------------------------------------------------------------
%\section*{Acknowledgments}
%-------------------------------------------------------------------------------

%The USENIX latex style is old and very tired, which is why there's no \textbackslash{}acks command for you to use when acknowledging. Sorry.

%-------------------------------------------------------------------------------

%-------------------------------------------------------------------------------
\clearpage
\bibliographystyle{plain}
\bibliography{reference}

%%%%%%%%%%%%%%%%%%%%%%%%%%%%%%%%%%%%%%t%%%%%%%%%%%%%%%%%%%%%%%%%%%%%%%%%%%%%%%%%%
\end{document}